\documentclass[12pt]{article}

\def\be{\begin{equation}}
\def\ee{\end{equation}}

\usepackage{graphicx,amsmath,amssymb}
\usepackage{multirow}
\usepackage{subfigure}

\begin{document}
\titlepage
\begin{flushright}
IPPP/16/01 \\
\today \\
\end{flushright}

\vspace*{0.5cm}
\begin{center}
{\Large \bf  The photon PDF in events with rapidity gaps}\\

\vspace*{1cm}

L. A. Harland-Lang$^a$, V. A. Khoze$^{b,c}$
and M. G. Ryskin$^c$

\vspace*{0.5cm}
$^a$ Department of Physics and Astronomy, University College London, WC1E 6BT, UK \\           
$^b$ Institute for Particle Physics Phenomenology, Durham University, DH1 3LE, UK    \\
$^c$
 Petersburg Nuclear Physics Institute, NRC Kurchatov Institute, 
Gatchina, St.~Petersburg, 188300, Russia \\
\end{center}

\begin{abstract}
\noindent We consider photon--initiated events with large rapidity gaps in proton--proton collisions,  where one or both protons may break up. We formulate a modified photon PDF that accounts for the specific experimental rapidity gap veto, and demonstrate how the soft survival probability for these gaps may be implemented consistently. Finally, we present some phenomenological results for the two--photon induced production of lepton and $W$ boson pairs.
\end{abstract}
\vspace*{0.5cm}

\section{Introduction}

Photon--initiated processes at the LHC allow us to study $\gamma p$ and two--photon interactions at unprecedented collision energies, for a range of final states.  In inclusive processes taking account of electroweak corrections is of increasing importance for precision phenomenology, and an essential ingredient in these is the introduction of a photon parton distribution function (PDF), where data such as the electroproduction of an isolated photon $ep \to e\gamma X$ at HERA, and electroweak boson production at the LHC are sensitive to the size of the photon distribution (see~\cite{Martin:2004dh,Ball:2013hta,Schmidt:2015zda} for studies by the global parton fitting groups). 

In addition to the inclusive case,  it also natural to consider photon--initiated exclusive and diffractive processes. The colour--singlet photon exchange can lead naturally to rapidity gaps in the final state, and in addition these modes offer some important and potentially unique advantages. For example, diffractive vector meson production provides a probe of the gluon PDF at low $x$ and $Q^2$, as well as possible gluon saturation effects, $\gamma\gamma\to W^+W^-$ pair production provides a precise probe of potential anomalous gauge couplings~\cite{Kepka:2008yx,Chapon:2009hh,Royon:2015coa}, while the theoretically well understood case of lepton pair production, $\gamma\gamma\to l^+l^-$, is sensitive to the effect of soft proton interactions~\cite{Aad:2015bwa,Harland-Lang:2015cta} as well as potentially being useful for  luminosity calibration~\cite{Khoze:2000db}. Moreover, there has recently been a renewal of interest in photon--induced processes in light of the excess of events at 750 GeV in the diphoton mass spectrum seen by ATLAS~\cite{ATLAS:2015exc} and CMS~\cite{CMS:2015dxe}; if this is due to a new resonance with a sizeable coupling to photons, then the $\gamma\gamma$--induced production mechanism may be dominant, see e.g.~\cite{Fichet:2015vvy,Csaki:2015vek,Fichet:2016pvq,Csaki:2016raa}.

Such processes may be purely exclusive, that is with the outgoing protons remaining intact after the collision, or the interacting protons may dissociate. In the latter case, while there will be additional secondary particle production in some rapidity interval, the events can nonetheless have a diffractive topology, and thus an attractive way to select photon exchange events is to require a Large Rapidity Gap (LRG) between the centrally produced system ($W^+W^-$ or $l^+l^-$ pair, $J/\psi$ or $\Upsilon$, etc) and the forward outgoing secondaries. There are a range of measurement possibilities for such processes at the LHC, with a promising experimental programme underway~\cite{yp}. The potential for rapidity gap vetoes to select events with a diffractive topology is in particular relevant at LHCb, for which the relatively low instantaneous luminosity and wide rapidity coverage allowed by the newly installed HERSCHEL forward detectors~\cite{Albrow:2014lta} are highly favourable, while similar scintillation counters are also installed at ALICE~\cite{Schicker:2014wvk} and CMS~\cite{Albrow:2014lta}. In addition, exclusive events may be selected at the LHC by tagging the outgoing intact protons using the approved AFP~\cite{CERN-LHCC-2011-012} and installed CT--PPS~\cite{Albrow:1753795} forward proton spectrometers, associated with the ATLAS and CMS central detectors, respectively, see also~\cite{yp,Royon:2015tfa}. 

In this paper we will consider $\gamma\gamma$--induced reactions where the outgoing protons may dissociate (see~\cite{Harland-Lang:2015cta} and references therein for discussion of the purely exclusive case), but with large rapidity gaps present between the produced object and the outgoing proton dissociation products. Provided the experimental rapidity veto region is large enough, the remaining contribution from non $\gamma\gamma$--initiated processes (e.g. standard Drell--Yan production) will be small, and can be suppressed with further cuts and subtracted using MC simulation, see for example~\cite{Aad:2015bwa,Chatrchyan:2013akv}. When considering these processes, there are two important effects that must be correctly accounted for. First, the secondaries produced during the DGLAP evolution of the photon PDF may populate the LRG. This means that in order to calculate the corresponding cross section we can not use the conventional inclusive PDF which describes the probability to find a photon in the proton, without any additional restrictions. Rather, we have to construct a new, modified, PDF where the evolution equation is supplemented by the condition that no $s$--channel partons are emitted in the LRG interval. Second, we have to include the probability that the gap will not be filled by secondaries produced by additional soft interactions of the colliding protons. This gap survival factor, $S^2$, can be calculated within a given model of soft hadronic interactions, see e.g.~\cite{Khoze:2014aca,Gotsman:2014pwa}; although this introduces an element of model--dependence in the corresponding predictions, these can be fairly well constrained by the requirement that they give a satisfactory description of high energy proton--proton scattering data such as the differential elastic proton cross section, $d\sigma_{el}/dt$, and the proton dissociation cross sections. 

The outline of this paper is as follows. In Section~\ref{sec:mod} we describe how a rapidity gap veto may be accounted for in a modified photon PDF.  In Section~\ref{sec:gap} we discuss the inclusion of the survival factor and demonstrate the effects this has on the $\gamma\gamma$ luminosity. In Section~\ref{sec:pred} we present numerical predictions for lepton and $W$ boson pair production at  $\sqrt{s}=13$ TeV. Finally, in Section~\ref{sec:conc} we conclude.

\section{The modified photon distribution}\label{sec:mod}
The photon PDF is given in terms of an input term $\gamma(x,Q^2_0)$ at the starting scale $Q_0$, and a term due to  photon emission from quarks during the DGLAP evolution up to the hard scale $\mu$. Since the QED coupling $\alpha$ is very small it is sufficient to consider just the leading $O(\alpha)$ contribution to the evolution, while the appropriate splitting functions which allow the evolution to be evaluated at NLO in the strong coupling $\alpha_S$ have recently been calculated in~\cite{deFlorian:2015ujt}, and are included here\footnote{In general, to be consistent  the NLO correction to the $\gamma\gamma \to X$ matrix element should also be included. However for the production of colourless particles that we will consider in this paper, these are zero.}. The photon PDF is thus given by
\begin{align}\nonumber
\gamma(x,\mu^2)&=\gamma(x,Q_0^2)+\int_{Q_0^2}^{\mu^2}\frac{\alpha(Q^2)}{2\pi}\frac{dQ^2}
{Q^2}\int_x^1\frac{dz}z \bigg(P_{\gamma\gamma}(z)\gamma(\frac xz,Q^2)\\ \label{pdf}
&+\sum_q e^2_qP_{\gamma q}(z)q(\frac xz,Q^2)+ P_{\gamma g}(z)g(\frac xz,Q^2)\bigg)\;,
\end{align}
where $\gamma(x,Q_0^2)$ is the input photon distribution at the scale $Q_0$. This may be written in terms of a coherent component due to the elastic process, $p\to p+\gamma$, and $p\to N^*+\gamma$ excitation, see~\cite{Martin:2014nqa}, as well as a component due to emission from the individual quarks within the proton (i.e. the direct analogue of perturbative emission in the QCD case). The $P_{\gamma q}(z)$ and $P_{\gamma g}(z)$ are the NLO (in $\alpha_S$) splitting functions. At LO we have
 \begin{align}
 P_{\gamma g}(z)&=0\; , \\
 P_{\gamma q}(z)&=\left[\frac{1+(1-z)^2}z\right]\;,\\
 P_{\gamma\gamma}(z)&=-\frac{2}{3}\left[N_c\sum_q e^2_q +\sum_l e^2_l\right]\delta(1-z)\;,
 \end{align}
 where the indices $q$ and $l$ denote the light quark and the lepton flavours respectively\footnote{In~\cite{Martin:2014nqa} the lepton contribution to $P_{\gamma\gamma}$ was mistakenly omitted.}, see~\cite{deFlorian:2015ujt} for the full NLO results. 
In fact, if we ignore the small corrections that the photon PDF will give to the evolution of the quark and gluons  then the equation (\ref{pdf}) for the DGLAP evolution of the photon PDF can be solved exactly, giving
\begin{align}\nonumber
\gamma(x,\mu^2)&=\gamma(x,Q_0^2)\,S_{\gamma}(Q_0^2,\mu^2)+\int_{Q_0^2}^{\mu^2}\frac{\alpha(Q^2)}{2\pi}\frac{dQ^2}
{Q^2}\int_x^1\frac{dz}z \bigg(\;\sum_q e^2_qP_{\gamma q}(z)q(\frac xz,Q^2)\\ \label{pdf1}
&+ P_{\gamma g}(z)g(\frac xz,Q^2)\bigg)\,S_{\gamma}(Q^2,\mu^2)\;,\\ \label{pdf1i}
&\equiv \gamma^{{\rm in}}(x,\mu^2)+\gamma^{\rm evol}(x,\mu^2)\;,
\end{align}
where the photon Sudakov factor
\be\label{sudgam}
S_{\gamma}(Q_0^2,\mu^2)=\exp\left(-\frac{1}{2}\int_{Q_0^2}^{\mu^2}\frac{{\rm d}Q^2}{Q^2}\frac{\alpha(Q^2)}{2\pi}\int_0^1 {\rm d} z\sum_{a=q,\,l}\,P_{a\gamma}(z)\right)\;,
\ee
corresponds to the probability for the photon PDF to evolve from scales $Q_0$ to $\mu$ without further branching; here $P_{q(l)\gamma}(z)$ is the $\gamma$ to quark (lepton) splitting function at NLO in $\alpha_s$. At LO it is given by
\be
P_{a\gamma}(z)=N_a\left[z+(1-z)^2\right]\;,
\ee
where $N_a=N_c e_q^2$ for quarks and $N_a=e_l^2$ for leptons, while the factor of $1/2$ in (\ref{sudgam}) is present to avoid double counting over the quark/anti--quarks (lepton/anti--leptons). The Sudakov factor is generated by resumming the term proportional to $P_{\gamma\gamma}$, due to virtual corrections to the photon propagator, which is a relatively small correction to the photon evolution. However this correction is not negligible, in particular for larger masses; we have  $S_{\gamma}\sim 0.97-0.93$ for $M_X=20-500$ GeV. 

As described above, the solution (\ref{pdf1}) is only exact if we neglect the dependence of the quark and gluon PDFs on the photon PDF, through the $P_{q\gamma}$ and $P_{g\gamma}$ terms in their evolution, respectively. These correspond to $O(\alpha^2)$ corrections to the photon evolution, and are therefore formally higher--order in $\alpha$, so that they can be safely neglected. To confirm this expectation, we have compared  (\ref{pdf1}) with the result of solving (\ref{pdf}) numerically with the $P_{q\gamma}$ term included in the quark evolution, at LO in $\alpha_S$ and only considering QED evolution (i.e. using the  \texttt{QECDS} scheme~\cite{Bertone:2013vaa} described below) for concreteness;  the contribution from $P_{g\gamma}$ only enters at NLO in $\alpha_S$ and so will be further suppressed. As expected, the difference is very small, and the results are found to coincide to within less than 0.1\%. We have also confirmed this by using the \texttt{APFEL} evolution code~\cite{Bertone:2013vaa}, with the results with and without the $P_{q\gamma}$ term in the quark evolution coinciding to a very similar level, irrespective of the evolution scheme used.

\begin{figure} [t]
\begin{center}
\includegraphics[scale=1.0]{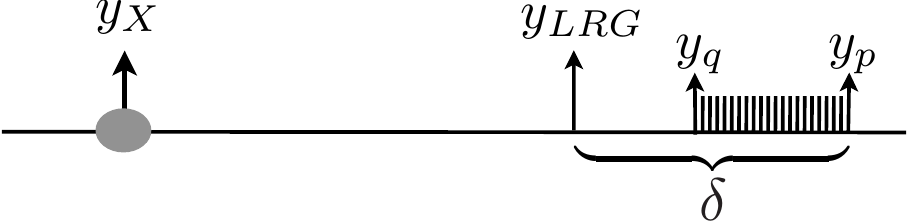}
\caption{Schematic diagram corresponding to the diffractive topology described in text, where a quark of rapidity $y_q$ is emitted beyond the edge of a LRG region.}
\label{fig:gap}
\end{center}
\end{figure}

The above equations correspond to the fully inclusive distribution, that is without any gap survival conditions. To include these, we note that as shown in (\ref{pdf1i}) the photon PDF at a scale $\mu$ may be expressed as a sum of a term, $\gamma^{\rm in}(x,\mu^2)$, due to the input PDF, i.e. generated by coherent and incoherent photon emission up to the scale $Q_0$, multiplied by the probability of no further emission up to the hard scale $\mu$, and a second term, $\gamma^{\rm evol}(x,\mu^2)$, due purely to DGLAP emission from the quark/gluons, which is independent of the input photon PDF.  For the coherent input component, there is naturally a large rapidity gap between the intact proton or the $N^*$ excitation and the central system. This is also the case for the incoherent input term: since the input value of $Q^2_0\sim 1$ ${\rm GeV}^2$ is small and we have the kinematic requirement on the quark $q_t<Q_0$, this implies that the transverse momentum of the final--state quark produced in the incoherent emission is small, and the rapidity of the produced secondaries is large, that is, similar to the outgoing proton/$N^*$ rapidity in the coherent case\footnote{For very large rapidity gaps, i.e. as the limit of the gap region approaches $y_{\rm max}=\pm\log(m_p/\sqrt{s})$, the decay products of the excited system may extend into the gap region. In this case, we have to consider each component of the input contribution individually, keeping only the part for which the  produced secondaries do not spoil the rapidity gap.}.

Next, we consider the second term in (\ref{pdf}), that is due to the DGLAP evolution. At LO, this corresponds to the splitting of a quark with a fraction $x/z$ of the proton momentum $\vec{p}$ to a photon with longitudinal momentum $xp$ (where $p=|\vec{p}|$) and squared transverse momentum 
\be\label{kt}
q_t^2=(1-z)\,Q^{2}\;,
\ee
and to a $s$--channel quark with longitudinal momentum 
\be\label{long}
p'_q=\frac{x(1-z)}{z}\,p\; . 
\ee
Due to strong $q_t$ ordering, the transverse momentum of the recoiled quark is given by $-q_t$, that is equal and opposite to that of the final--state photon.
The rapidity of this quark is
\be
y_q\simeq -\ln\frac{q_t}{2p'_q}\ .
\label{eta}
\ee 
We require that the quark be produced with rapidity greater than some $y_{LRG}$, corresponding to the end of the experimentally defined gap\footnote{For consistency we work in terms of particle rapidities, although experimentally it is generally the pseudorapidity $\eta_{\rm LRG}=-\ln\left[\tan\left(\theta_{\rm LRG}/2\right)\right]$ which defines the edge of the gap; for massless particles these variables are of course equivalent.}: in this case, it is convenient to work in terms of the rapidity interval, $\delta=y_p-y_{LRG}$ between the edge of the gap and outgoing proton in which the quark may be emitted, see Fig.~\ref{fig:gap}. The condition $y_q >y_{LRG}$ in this notation takes the form
\be
y_p-y_q=\ln\left(\frac{q_t}{m_p}\frac {z} {x(1-z)}\right) < \delta \ ,
\label{dif}
\ee 
and thus to obtain the modified photon PDF, corresponding to the kinematics with a LRG present, we have to simply supplement the integrand in (\ref{pdf1}) by a $\Theta$ function which ensures that the condition (\ref{dif}) is satisfied. This gives\footnote{If we consider the evolution equation (\ref{pdf1}) in terms of the scale $q_t$ then the limit (\ref{dif}) corresponds to a straightforward upper limit on the momentum fraction, $z$.}
\begin{align}\nonumber
\gamma(x,\mu^2)&=\gamma(x,Q_0^2)\,S_{\gamma}(Q_0^2,\mu^2)+\int_{Q_0^2}^{\mu^2}\frac{\alpha(Q^2)}{2\pi}\frac{{\rm d}Q^2}
{Q^2}\int_x^1\frac{dz}z \bigg(\;\sum_q e^2_qP_{\gamma q}(z)q(\frac xz,Q^2)\\ \label{pdfm}
&+ P_{\gamma g}(z)g(\frac xz,Q^2)\bigg)\,S_{\gamma}(Q^2,\mu^2) \Theta\left[e^\delta-\frac{q_t}{m_p}\frac{z}{x(1-z)}\right]\;,\\ \label{photdef1}
&\equiv \gamma^{{\rm in}}(x,\mu^2)+\gamma^{\rm evol}(x,\mu^2;\delta)\;,
\end{align}
where $q_t=\sqrt{(1-z)Q^{2}}$ and in the final expression serves to define the $\delta$--dependent evolution component $\gamma_{\rm evol}(x,\mu^2;\delta)$. Due to strong $q_t$ ordering all the previous partons emitted during the evolution will have larger rapidities, $y>y_q$, and therefore evidently do not spoil the rapidity gap; this condition is therefore sufficient for a LRG to be present.

We now consider some numerical results. As described above, for the input photon PDF, following~\cite{Martin:2014nqa} we include a coherent component due to purely elastic photon emission and an incoherent component due to emission from the individual quark lines, such that
\begin{equation}\label{inputdef}
\gamma(x,Q_0^2)=\gamma_{\rm coh}(x,Q_0^2)+\gamma_{\rm incoh}(x,Q_0^2)\;,
\end{equation}
with
\begin{equation}\label{gamcoh}
\gamma_{\rm coh}(x,Q_0^2)=\frac{1}{x}\frac{\alpha}{\pi}\int_0^{Q^2<Q_0^2}\!\!\frac{{\rm d}q_t^2 }{q_t^2+x^2 m_p^2}\left(\frac{q_t^2}{q_t^2+x^2 m_p^2}(1-x)F_E(Q^2)+\frac{x^2}{2}F_M(Q^2)\right)\;,
\end{equation}
where  $q_t$ is the transverse momentum of the emitted photon, and $Q^2$ is the modulus of the photon virtuality, given by
\begin{equation}\label{tdef}
Q^2=\frac{q_t^2+x^2m_p^2}{1-x}\;,
\end{equation}
The functions $F_E$ and $F_M$ are the usual proton electric and magnetic form factors
\begin{equation}\label{form1}
F_M(Q^2)=G_M^2(Q^2)\qquad F_E(Q^2)=\frac{4m_p^2 G_E^2(Q^2)+Q^2 G_M^2(Q^2)}{4m_p^2+Q^2}\;,
\end{equation}
with
\begin{equation}\label{form2}
G_E^2(Q^2)=\frac{G_M^2(Q^2)}{7.78}=\frac{1}{\left(1+Q^2/0.71 {\rm GeV}^2\right)^4}\;,
\end{equation}
in the dipole approximation, where $G_E$ and $G_M$ are the `Sachs' form factors. The incoherent input term is given by\footnote{In fact, we take the slightly different form described in footnote 3 of~\cite{Martin:2014nqa}, with as in (\ref{gamincoh}) the replacement $F_1(Q^2)\to G_E(Q^2)$ made to give a more precise evaluation for the probability of coherent emission.}
\begin{equation}\label{gamincoh}
\gamma_{\rm incoh}(x,Q_0^2)=\frac{\alpha}{2\pi}\int_x^1\frac{{\rm d}z}{z}\left[\frac{4}{9}u_0\left(\frac{x}{z}\right)+\frac{1}{9}d_0\left(\frac{x}{z}\right)\right]\frac{1+(1-z)^2}{z}\int^{Q_0^2}_{Q^2_{\rm min}}\frac{{\rm d}Q^2}{Q^2+m_q^2}\left(1-G_E^2(Q^2)\right)\;,
\end{equation}
where
\begin{equation}
Q^2_{\rm min}=\frac{x}{1-x}\left(m_\Delta^2-(1-x)m_p^2\right)\;,
\end{equation}
accounts for the fact that the lowest proton excitation is the $\Delta$--isobar, and the final factor $(1-G_E^2(Q^2))$ corresponds to the probability to have no intact proton in the final state (which is already included in the coherent component). Here $m_q=m_d$($m_u$) when convoluted with $d_0$($u_0$), and the current quark masses are taken. As the quark distributions are frozen for $Q<Q_0$, this represents an upper bound on the incoherent contribution. Although other models for this incoherent component may also be taken, the conclusions which follow are relatively insensitive to the specific choice, and so for simplicity we will not consider them here. We also note that it is possible to account explicitly for the first $\Delta$--isobar excitation in the coherent component, see~\cite{Martin:2014nqa}, however this does not have a noticeable effect on the results which follow, and is omitted here.

\begin{figure}
\begin{center}
\includegraphics[scale=0.65]{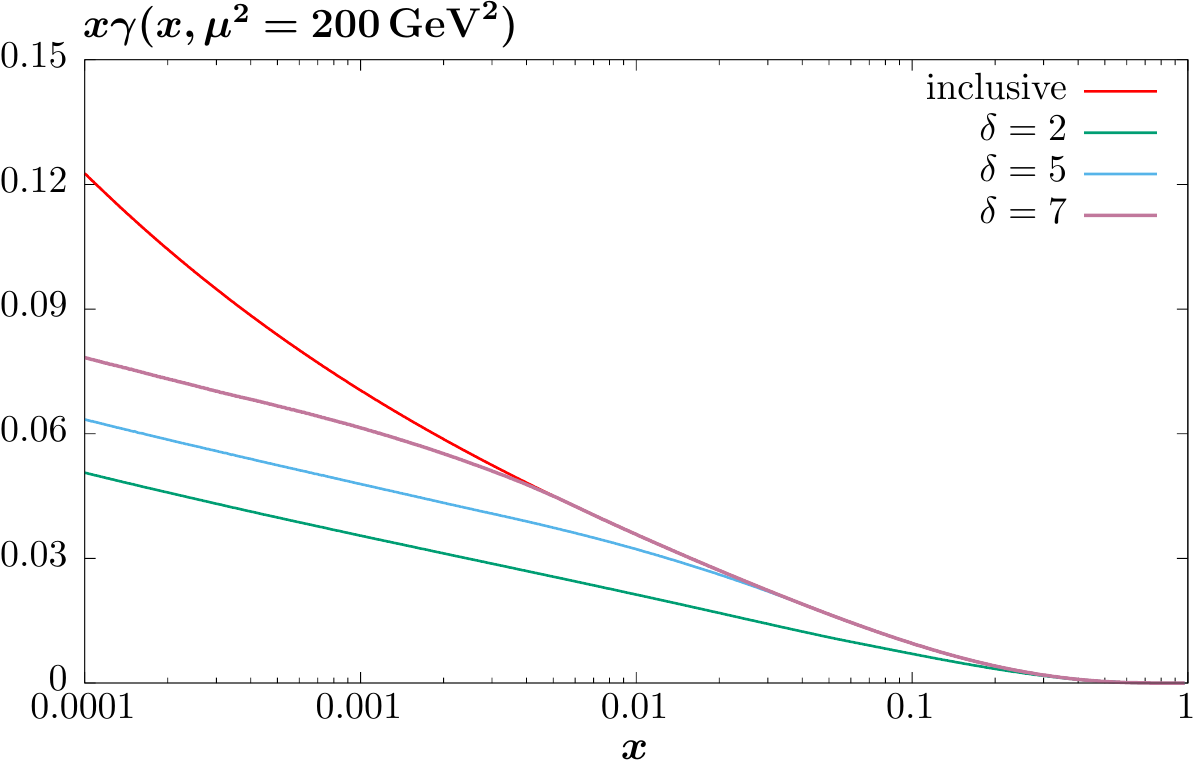}
\includegraphics[scale=0.65]{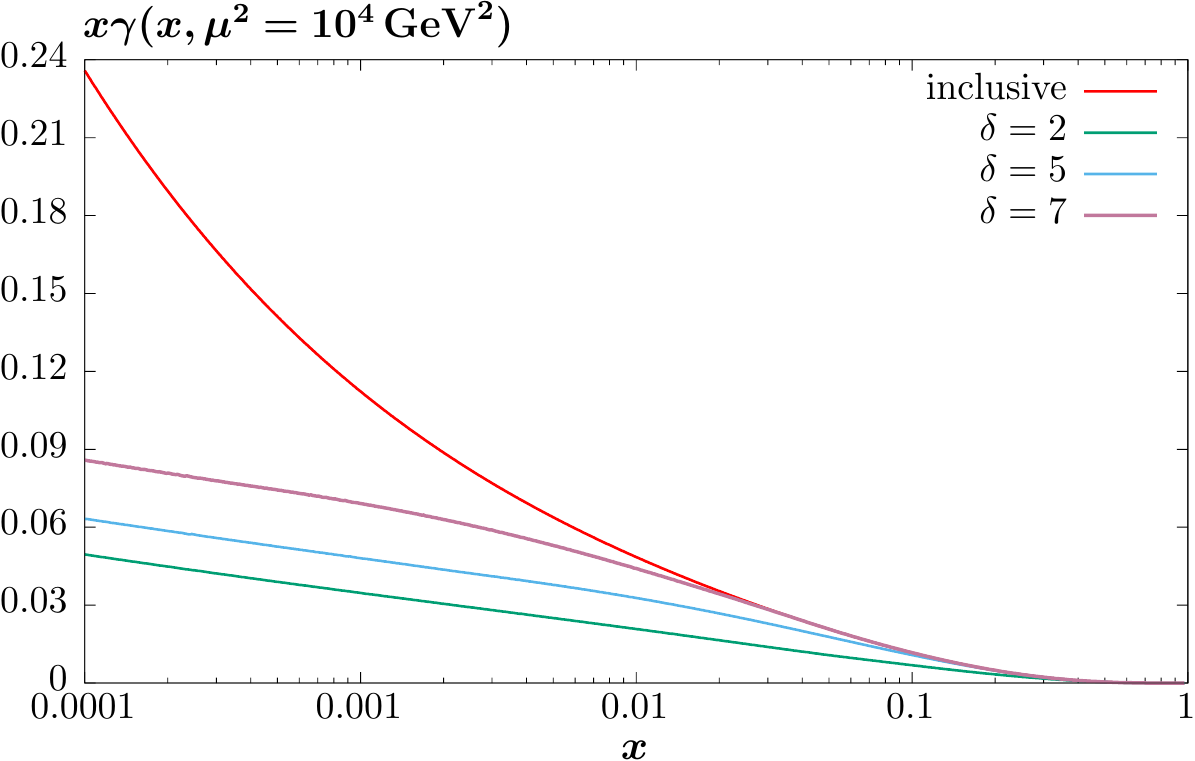}
\caption{The photon PDF $x\gamma(x,\mu^2)$ subject to the rapidity gap constraint (\ref{dif}), for different values of $\delta$ and for $\mu^2=200, \, 10^4\,{\rm GeV}^2$, with the usual inclusive PDF shown for comparison.}
\label{fig:split}
\end{center}
\end{figure}

\begin{figure}
\begin{center}
\includegraphics[scale=0.65]{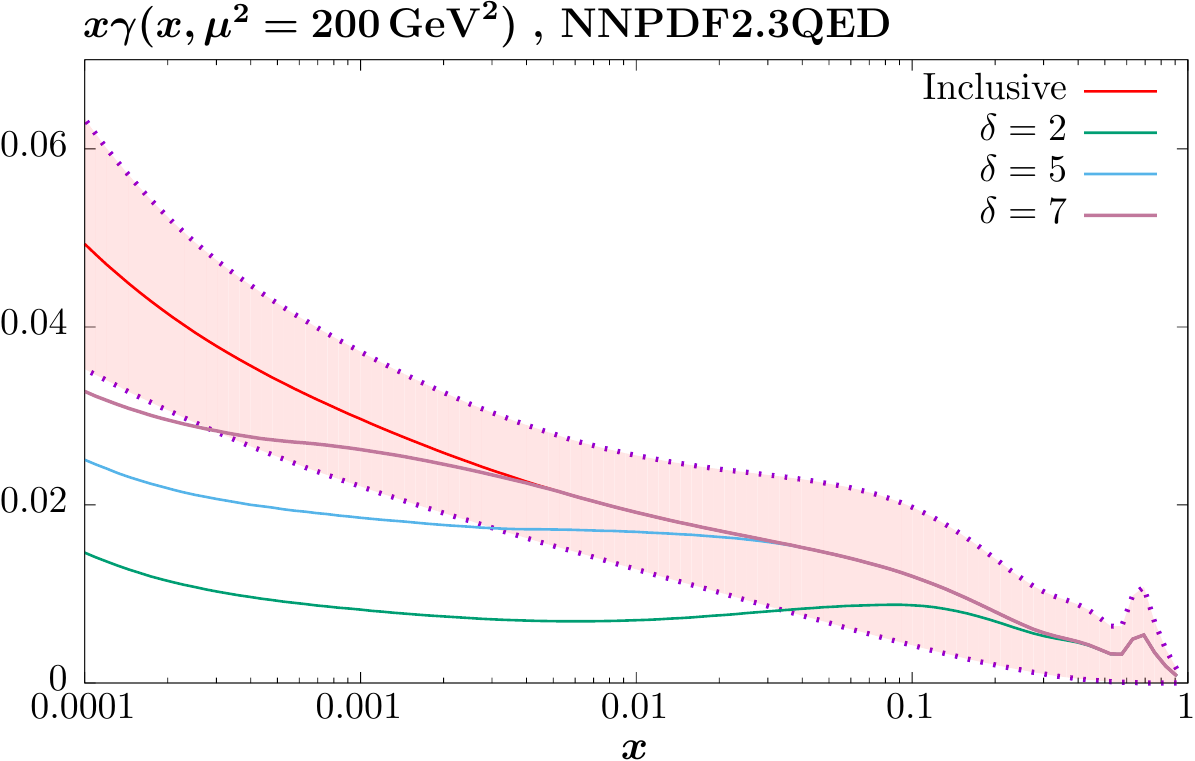}
\includegraphics[scale=0.65]{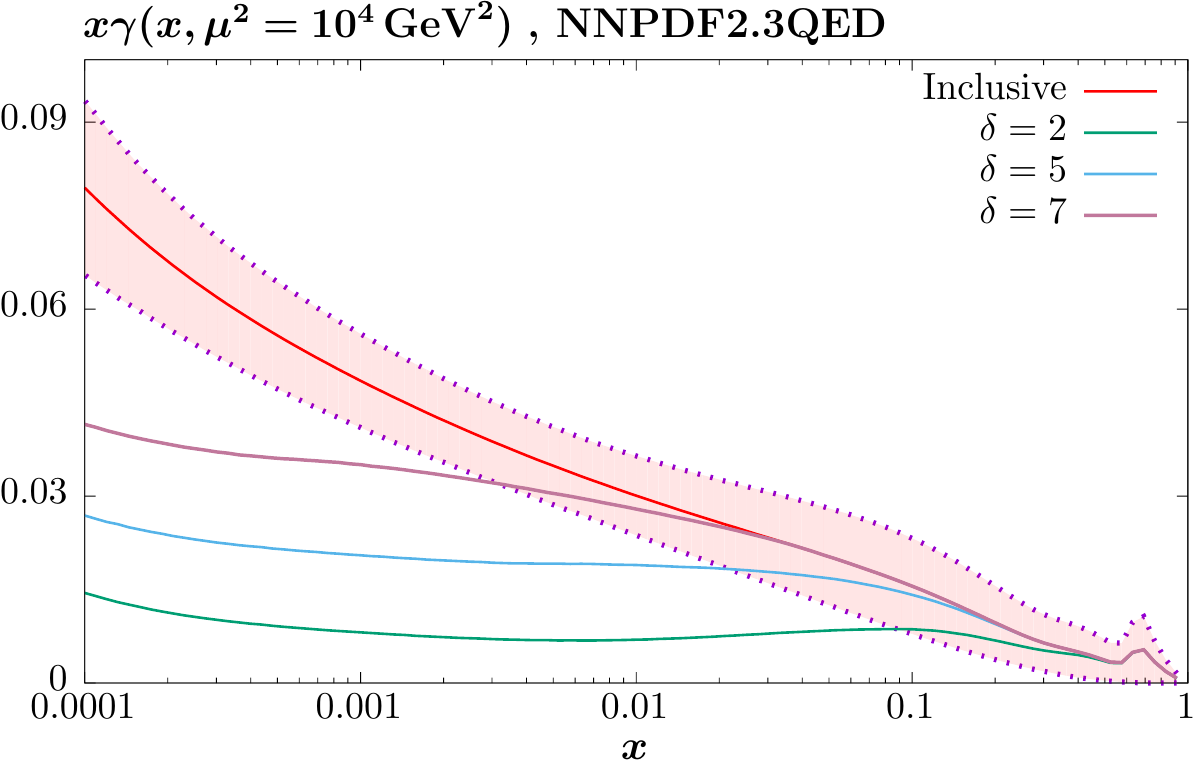}
\caption{As in Fig.~\ref{fig:split}, but with the NNPDF2.3QED~\cite{Ball:2013hta} set taken for the input PDF at $Q_0^2=2\,{\rm GeV}^2$. The 68\% confidence error bands are shown in the inclusive case.}
\label{fig:splitnn}
\end{center}
\end{figure}

In Fig.~\ref{fig:split} we show the effect of including the rapidity gap constraint (\ref{dif}) on the photon PDF, for two choice of scale and for different values of $\delta$. Here, and in all numerical results which follow, we for concreteness use MMHT2014 NLO PDFs~\cite{Harland-Lang:2014zoa} for the quark term in (\ref{pdfm}). The suppression in the PDFs relative to the inclusive case, which becomes stronger as  $\delta$ decreases, is clear. In addition, we can see that the suppression is stronger at lower $x$ and higher $\mu^2$, as expected from (\ref{dif}): in the former case, the outgoing quark in the $q\to q\gamma$ splitting has on average lower longitudinal momentum, while in the latter the quark transverse momentum is higher, such that in both cases the quark tends to be produced more centrally. These effects are not limited to the particular approach to modelling the photon PDF described above: in Fig.~\ref{fig:splitnn} we show the same plots as before, but using the NNPDF2.3QED~\cite{Ball:2013hta} photon PDF as input\footnote{We note that the PDF evolution for the NNPDF set is performed in the so--called \texttt{QECDS} scheme~\cite{Bertone:2013vaa}, where the QED and QCD factorization scales are treated seperately, with the QED evolution performed first. In the context of our approach, this corresponds to evaluating the quark PDFs in (\ref{pdf}) at fixed scale $Q_0^2$ (we treat the QED evolution here at LO in $\alpha_s$, consistently with NNPDF, and hence no gluon term is present). However, as discussed in~\cite{Bertone:2013vaa}, this \texttt{QECDS} scheme leads to potentially large unresummed logarithms at higher scales, and we use it here and in all NNPDF results which follow only for the sake of comparison.}. The increased suppression with decreasing $x$ and increasing $Q^2$ is again clear, with the resulting PDFs generally lying outside the uncertainty band in the inclusive PDF. 

We end this section with some comments. First, we note that qualitatively speaking the inclusion of the $\Theta$ function in the integral (\ref{pdfm}) plays the role of the Sudakov factor in gluon--mediated central exclusive production (CEP) processes, see e.g.~\cite{Harland-Lang:2014lxa}, that is, it accounts for the probability for no  secondary partons emission. In the case of pure CEP processes, such emission is entirely forbidden, whereas here we only require that no secondaries are emitted into the veto region. Second, in accounting for the veto condition (\ref{dif}) in the case of the NLO splitting functions we should consider vetoes on the two emitted partons individually, i.e. $qg$($q\overline{q}$) for  $P_{\gamma q}$($P_{\gamma g}$). However since the effect of the NLO correction is rather small ($\sim 5$\% ) here we for simplicity use the same veto as in the LO case. This corresponds to a veto on the kinematics of the parton pair and so only gives an approximate indication of the effect to the NLO contribution. In addition, we emphasise that (\ref{pdfm}) corresponds to the survival of the LRG in terms of the secondary partons only. A complete evaluation, in which the probability that no secondary  hadrons spoil the gap would require a Monte Carlo simulation which accounts for the fluctuations during the fragmentation and hadronization process. However, the results presented above should not be too sensitive to these effects, in particular if $\delta$ is large enough. Finally, we note that the photon PDF is formally defined for fully inclusive processes, i.e. where a complete sum over all final states is performed. For the distribution in the semi--exclusive case,  only a subset of final states is summed over, and so the PDF no longer has this formal definition. Moreover, as discussed in the following section, factorization is explicitly violated by soft survival effects. Nonetheless, we may consider these PDFs as phenomenological objects which capture the important physics of the semi--exclusive process, i.e. the constraint (\ref{dif}), for which the $Q^2$--evolution can be described within the same leading logarithmic approximation as in the standard DGLAP approach.

\section{Soft gap survival factor}\label{sec:gap}

In addition to accounting for the rapidity gap veto in the $q\to q\gamma$ splitting associated with the DGLAP evolution of the photon PDF, we must also consider the possibility of additional soft proton--proton rescattering, that is underlying event activity, which can fill the rapidity gaps with secondary particles. The probability that this does not occur is encoded in the eikonal survival factor, $S^2_{\rm elk}$, see~\cite{Khoze:2014aca,Gotsman:2014pwa} for some more recent theoretical work, and~\cite{Harland-Lang:2015cta,Harland-Lang:2014lxa}  for further discussion and references\footnote{Due to the strong $q_t$ ordering in DGLAP evolution we can safely neglect the effect of the so--called `enhanced' survival factor, see e.g.~\cite{Ryskin:2011qe}, generated by additional interactions with the intermediate partons produced during the evolution.}.

Here we follow the approach described in detail in~\cite{Harland-Lang:2015cta}, where it is emphasised that the impact of survival effects depends sensitively on the subprocess, through the specific proton impact parameter dependence. For example, for exclusive photon exchange processes, the low virtuality  (and hence transverse momentum) of the quasi--real photon exchange, which corresponds to relatively large impact parameters between the colliding protons, leads to an average survival factor that is close to unity, while for the less peripheral QCD--induced exclusive processes the suppression is much larger. This fact has important consequences in the current situation. In particular,  the expression (\ref{pdfm}) for the photon PDF evaluated at a scale $\mu^2$ is given in terms of an input distribution corresponding to low photon $q_\perp^2 < Q_0^2$, and a term generated by DGLAP evolution, for which we have $q_\perp^2 \gg Q_0^2$. This larger scale (and hence smaller average impact parameter)  suggests that the survival factor in the latter case will be much smaller.

\begin{figure}
\begin{center}
\subfigure[bare]{
\includegraphics[scale=1.3]{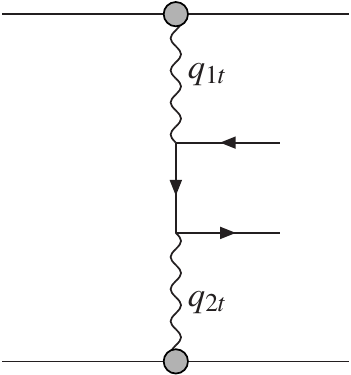}
}\qquad\qquad
\subfigure[screened]{
\includegraphics[scale=1.3]{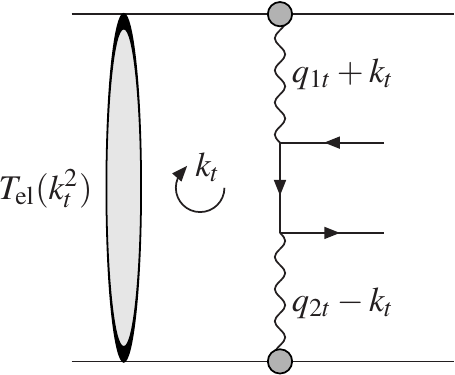}
}
\caption{Feynman diagrams for (a) bare and (b) screened amplitudes for coherent photon--induced lepton pair production}
\label{fig:survcoh}
\end{center}
\end{figure}

To demonstrate this, we must determine precisely how survival effects are to be included in this semi--exclusive case. First, we recall how such corrections are included in the exclusive case, relevant for the input component of the photon PDF. Here, we can work at the amplitude level, as described in detail in~\cite{Harland-Lang:2015cta}, where purely exclusive two--photon initiated processes are considered. Survival effects are generated by the so--called `screened' amplitude shown in Fig.~\ref{fig:survcoh}, for the representative case of lepton pair production, where the grey oval represents the exchange of a pomeron with transverse momentum $k_t$ transferred through the loop. The amplitude including rescattering corrections is given by
\begin{equation}\label{skt}
T^{\rm res}({q}_{1t},{q}_{2t}) = \frac{i}{s} \int\frac{{\rm d}^2  {k}_t}{8\pi^2} \;T_{\rm el}(k_t^2) \;T({q'}_{1t},{q'}_{2t})\;,
\end{equation}
where $q'_{1t}=q_{1t}+k_t$ and $q'_{2t}=q_{1t}-k_t$ are the incoming photon transverse momenta, as show in Fig.~\ref{fig:survcoh} (b), and other kinematic arguments of the amplitudes are omitted for simplicity. Here, $T(q'_{1t},q'_{2t})$ is the production amplitude; for $q'_{it}=q_{it}$ this corresponds to Fig.~\ref{fig:survcoh} (a). While the equivalent photon approximation and the corresponding expression (\ref{gamcoh}) for the coherent component of the photon PDF are defined at the cross section level, as discussed in~\cite{Harland-Lang:2015cta} the coherent amplitude may be unambiguously defined for the term proportional to the proton electromagnetic form factor, and thus included inside the $k_t$ integral (\ref{skt}). After adding to the `bare' amplitude (i.e. without survival effects) and squaring, the average survival factor may be evaluated using
\begin{equation}\label{seikav}
\langle S_{\rm eik}^2\rangle= \frac{\int {\rm d}^2 q_{1t}\,{\rm d}^2 q_{2t}\,|T(q_{1t},q_{2t})+T^{\rm res}(q_{1t},q_{2t})|^2}{\int {\rm d}^2q_{1t}\,{\rm d}^2q_{2t}\,|T(q_{1t},q_{2t})|^2}\;,
\end{equation}
where for illustration we have considered only the simplest, so--called  `one--channel'  approach, which ignores any internal structure of the proton: see~\cite{Ryskin:2009tk,Martin:2009ku} for discussion of how this can be generalized to the more realistic `multi--channel' case.

As discussed in~\cite{Harland-Lang:2015cta}, the inclusion of survival effects requires a non--trivial vector combination of the incoming photon transverse momenta $q_{it}$, which only after squaring and angular averaging allows the $q_{it}$ dependence to be factorized as in (\ref{gamcoh}). For example, for the production of a spin--0 object the production amplitude in (\ref{skt}) should be decomposed as
\be\label{agen}
T(q_{1t},q_{2t}) \sim -\frac{1}{2} ({\bf q}_{1t}\cdot {\bf q}_{2t})\,(T_{++}+T_{--})-\frac{i}{2} ({\bf q}_{1t}\times {\bf q}_{2t})\cdot {\bf n}_0\,(T_{++}-T_{--})
\ee
where the $T_{\lambda_1\lambda_2}$ are now the $\gamma(\lambda_1)\gamma(\lambda_2)\to X$ helicity amplitudes and ${\bf n}_0$ is a unit vector along the (beam) $z$--axis; see~\cite{Harland-Lang:2015cta} for the full case, including the $|J_z|=2$ amplitudes. Using this, it can readily be shown that the bare amplitude squared reduces to the correct cross section level expressions corresponding to (\ref{gamcoh}) or (\ref{gamincoh}), while in the screened case it is again crucial to include this correct vector form of the amplitude, i.e. with the ${q'}_{it}$ included inside the integral (\ref{skt}). We therefore do this here, following the approach of~\cite{Harland-Lang:2015cta}, for both the coherent and incoherent contributions. As the relative contributions of the  amplitudes $T_{\lambda_1\lambda_2}$ affect the $q_\perp$ dependence in (\ref{agen}), the survival factor will in general depend on the helicity structure of the $\gamma\gamma\to X$ process. In all numerical calculations that follow we consider for simplicity the case of a scalar object ($T_{++}=T_{--}$, $T_{\pm \mp}=0$). For other quantum numbers, the survival factor can vary by up to $\sim 10\%$.

\begin{figure}[h]
\begin{center}
\includegraphics[scale=1.0]{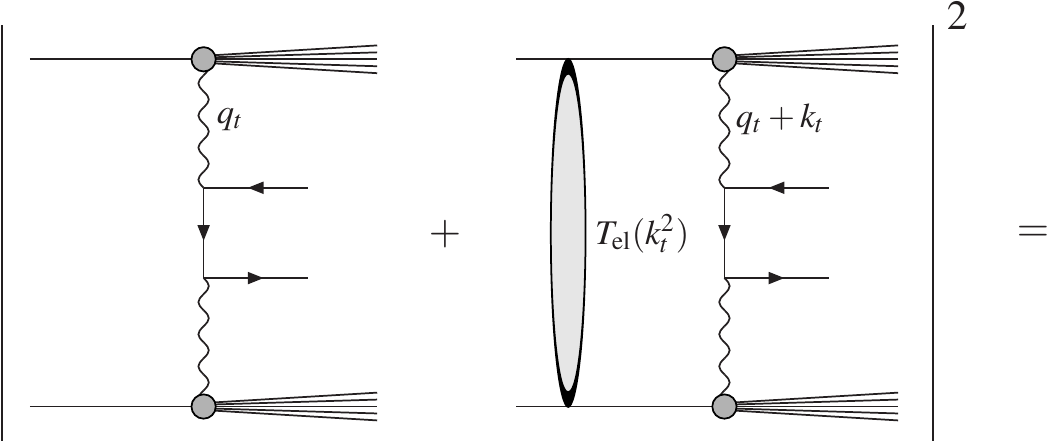}\par
\vspace{0.5cm}
\includegraphics[scale=0.55]{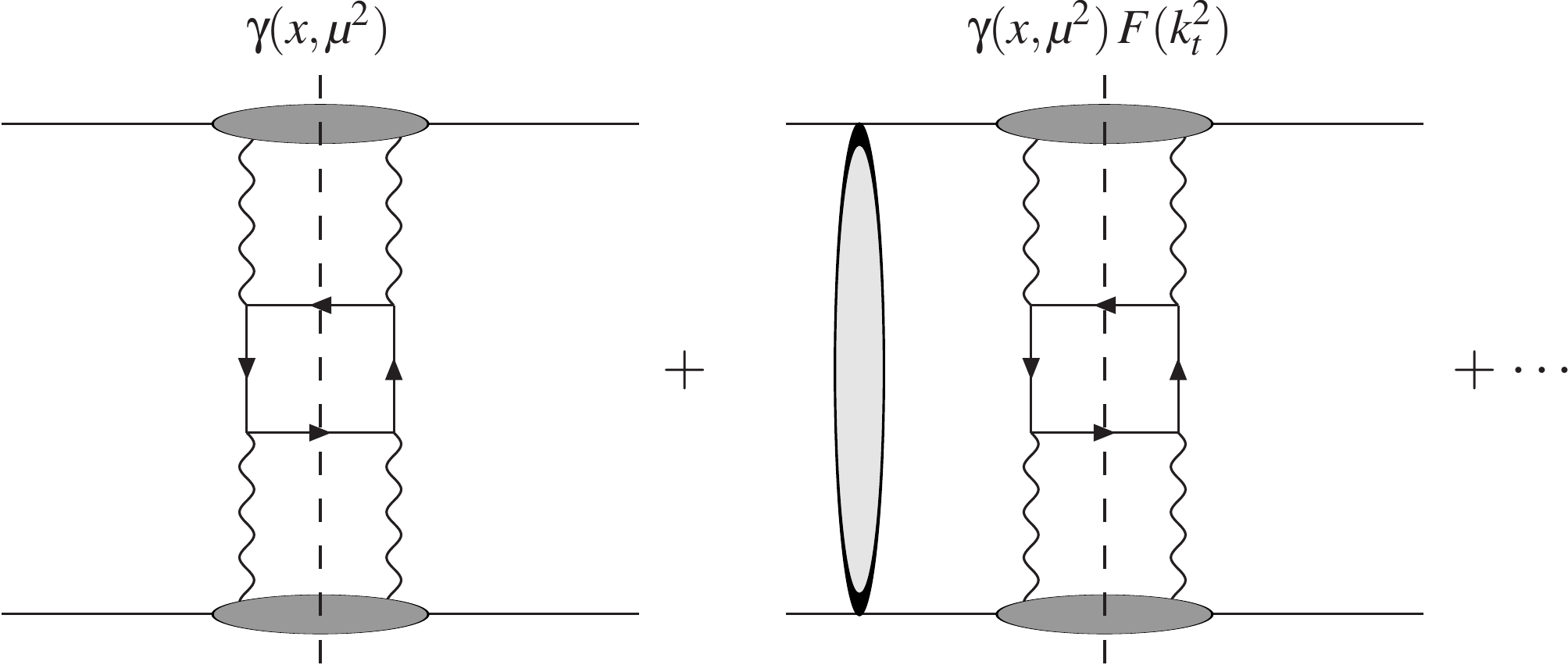}
\caption{Schematic Feynman diagrams indicating how screening effects may be included in the case of semi--exclusive photon--induced lepton pair production. The vertical dashed lines in the lower plots indicate the amplitude and the complex conjugate on the left and right hand side, respectively, while the remaining two diagrams with the pomeron exchange included in the conjugate amplitude and in both amplitude and the conjugate are not shown.}
\label{fig:surv}
\end{center}
\end{figure}

Next, we must consider the component of the photon PDF due to the DGLAP evolution term in (\ref{pdfm}), generated by the $q\to q\gamma$ splitting. We will first consider the case that the photon PDFs from both protons correspond to this evolution component, before discussing the mixed case. As mentioned above, in contrast to the input component, for which $q_t^2 < Q_0^2$,  we now have $q_t^2\gg Q_0^2$, due to strong DGLAP $q_t$ ordering. The transverse momentum transferred through the pomeron loop corresponds to a soft physics scale, and is generally very low, being set by $k_t^2 \sim 2/ B_{\rm el}$, where $B_{\rm el}$ is the $t$--slope for elastic $pp$ scattering, with at the LHC $B_{\rm el}\sim 20\,{\rm GeV}^{-2}$~\cite{Antchev:2013gaa} and so $k_t^2\sim 0.1 \, {\rm GeV}^2$. Thus, when we consider a screened diagram of the type shown in Fig.~\ref{fig:surv} (top right), we have $q_{it} \gg k_t $, and so when considering the screened amplitude (\ref{skt}), the incoming photon transverse momenta are simply given by $q'_{it}\sim q_{it}$, and the production amplitude $T(q'_{1t},q'_{2t})$ factorizes from the $k_t$ integral. 

More precisely, we note that in fact the evolution component can no longer be defined at the amplitude level, as it is the cross section which is written in terms of the photon PDF. This is shown schematically in Fig.~\ref{fig:surv} (bottom left) for the unscreened case. We must therefore define screening corrections at the cross section level; this can be achieved by observing that for the screened contribution to the cross section of the type shown in Fig.~\ref{fig:surv} (bottom left), instead of dealing with the usual `diagonal' photon PDF, we should instead consider the generalised PDF (GPDF), see~\cite{Diehl:2003ny,Belitsky:2005qn} for reviews and references, that is
\begin{equation}
x\gamma(x,\mu^2) \to H^\gamma(x,\xi=0,t;\mu^2)\;,
\end{equation}
where the skewedness is provided by the non--zero squared transverse momentum $t\approx -k_t$ transferred through the $t$--channel exchange (i.e. the hard process and the DGLAP ladder generating the evolution of the GPDF) by the screening pomeron\footnote{For the unscreened contribution in Fig.~\ref{fig:surv} (bottom left), due to the basics properties of the GPDFs we have $H^\gamma(x,\xi=0,t=0;\mu^2)=x\gamma(x,\mu^2)$, reproducing the usual factorisation formula, as it must do.}. We can then to good approximation neglect this $k_t$ dependence in the evolution of the GPDF: this is well justified at LO as due to the strong $q_t$ ordering of the DGLAP evolution, the $k_t \lesssim Q_0$ transferred through the pomeron loop can be neglected in every rung of the evolution ladder for $H^\gamma$ except that nearest the proton. This allows us to write the above expression in the factorized form
\begin{equation}\label{hft}
H^\gamma(x,\xi=0,t;\mu^2)=x\gamma^{\rm evol}(x,\mu^2;\delta)\,F_1(t)\;,
\end{equation}
where $\gamma^{\rm evol}$ is defined in (\ref{photdef1})  and $F_1$ is the proton Dirac form factor; this choice is motivated by the sum rule derived in~\cite{Ji:1996ek} for the quark GPDF, which is appropriate here as it is the quark PDF which is driving the photon evolution. This allows the $k_t$ dependence induced by the screening corrections to be properly accounted for at the cross section level, namely by expanding the squared amplitude in the numerator of (\ref{seikav}), giving
\begin{equation}\label{sigmascr}
\sigma^{\rm scr.}\sim 1 + \frac{2i}{s}\int\frac{{\rm d}^2 k_t}{8\pi^2}\,T_{\rm el}(k_t^2)F^2_1(k_t^2)-\frac{1}{s^2}\int\frac{{\rm d}^2 k_t}{8\pi^2}\frac{{\rm d}^2 k'_t}{8\pi^2}\,T_{\rm el}(k_t^2)T_{\rm el}({k'}_t^2)F^2_1((k_t+k'_t)^2)\;,
\end{equation}
where overall factors have been omitted for simplicity. This expansion is shown schematically in Fig.~\ref{fig:surv} (bottom). In fact, the calculation may be more easily performed in impact parameter space, in which case we simply have
\begin{equation}
\langle S^2_{\rm eik} \rangle =\int {\rm d}^2 b_t \, F^2_1(b_t) e^{-\Omega(b_t)}\;,
\end{equation}
where $b_t$ is the impact parameter separation of the two protons, $F_1(b_t)$ is the Fourier transform of the Dirac form factor, and $\Omega(b_t)$ is the so--called proton opacity; physically, $\exp\left(-\Omega(b_t)\right)$ represents the probability of no inelastic scattering at impact parameter $b_t$. Here, as above, we work in the single--channel approximation for the sake of clarity, but in actual calculations we use the two--channel approach described in~\cite{Khoze:2013dha}.

\begin{table}
\begin{center}
\begin{tabular}{|c|c|c|}
\hline
$\langle S^2\rangle$& $M_X^2=200 \,{\rm GeV}^2$ & $M_X^2=10^4 \,{\rm GeV}^2$  \\ \hline
(coh., coh.)& 0.95 & 0.89 \\
(coh., incoh.) &0.84  & 0.76\\
(incoh., incoh.)  &0.18  & 0.18 \\
(evol., coh.) &0.83  & 0.74 \\
(evol., incoh.) &0.16  & 0.16\\
(evol., evol.) &0.097  &0.097 \\
\hline
\end{tabular}
\caption{Average survival factor for different components of the photon PDFs $\gamma(x,M_X^2)$, for different systems of mass $M_X$ produced with $Y_X=0$. The coherent, incoherent and evolution components are shown for (proton 1, proton 2).}
\label{table:surv}
\end{center}
\end{table}

The above discussion only applies for the case that the evolution components are probed from both protons, however we must also consider the mixed case, where we probe the evolution component from one proton and the input from the other. Such a process will induce a distinct proton impact parameter distribution from the cases above, leading to a different survival factor. The precise size of this can be readily calculated by adjusting (\ref{sigmascr}) to include an integral over the photon $q_t$ emitted from the input side. However here care must again be taken as in (\ref{agen}) to include the correct vector dependence on the incoming photon $q_t$ on the input side. As the transverse momentum $q_{2t}$ (say) on the evolution side is much larger than the $k_t$ transferred through the exchanged pomeron, the $q_{2t}$ dependence factorizes from the $k_t$ integral in (\ref{skt}), and $q_{2t}$ can be averaged over at the cross section level, leaving only a dependence on $q_{1t}$ in the screened amplitude. We find that this averaging washes out the dependence (\ref{agen}) seen in the purely coherent/incoherent contributions, such that the survival factor here does not in fact depend on the quantum numbers of the produced object.

We now consider some numerical results for the average survival factors. To calculate these, we use model 4 of~\cite{Khoze:2013dha}, which applies a two--channel eikonal approach, in which the incoming proton is considered to be a coherent superposition of two diffractive `Good--Walker' (GW) eigenstates~\cite{Good:1960ba}, each of which may scatter elastically. In this situation, we have some freedom as to how the expression (\ref{hft}) may be suitably generalised, that is how the photon GPDF couples to the individual GW eigenstates. By default, we assume that these are the same as the coupling to the pomeron taken in~\cite{Khoze:2013dha}, that is $H^\gamma_i \sim \gamma_i \cdot F_1(t)$, for eigenstate $i=1, 2$, where $\gamma_i$ is coupling to the pomeron. However, this is not the only possibility: for larger $x$ where the quark contribution to $H^\gamma$ is more important, it may be more sensible to instead assume that this coupling is universal, i.e. simply $H^\gamma_i \sim  F_1(t)$.  A further question is whether the proton  form factor $F_1$ is the appropriate choice: it may be be more suitable, in particular at low $x$, to take the same form factors as in~\cite{Khoze:2013dha} for the coupling of the pomeron to the GW eigenstates.  In fact, it turns out that these different choices generally have a small effect on the observable predictions; we will comment on this further below.

 \begin{figure}
\begin{center}
\includegraphics[scale=0.65]{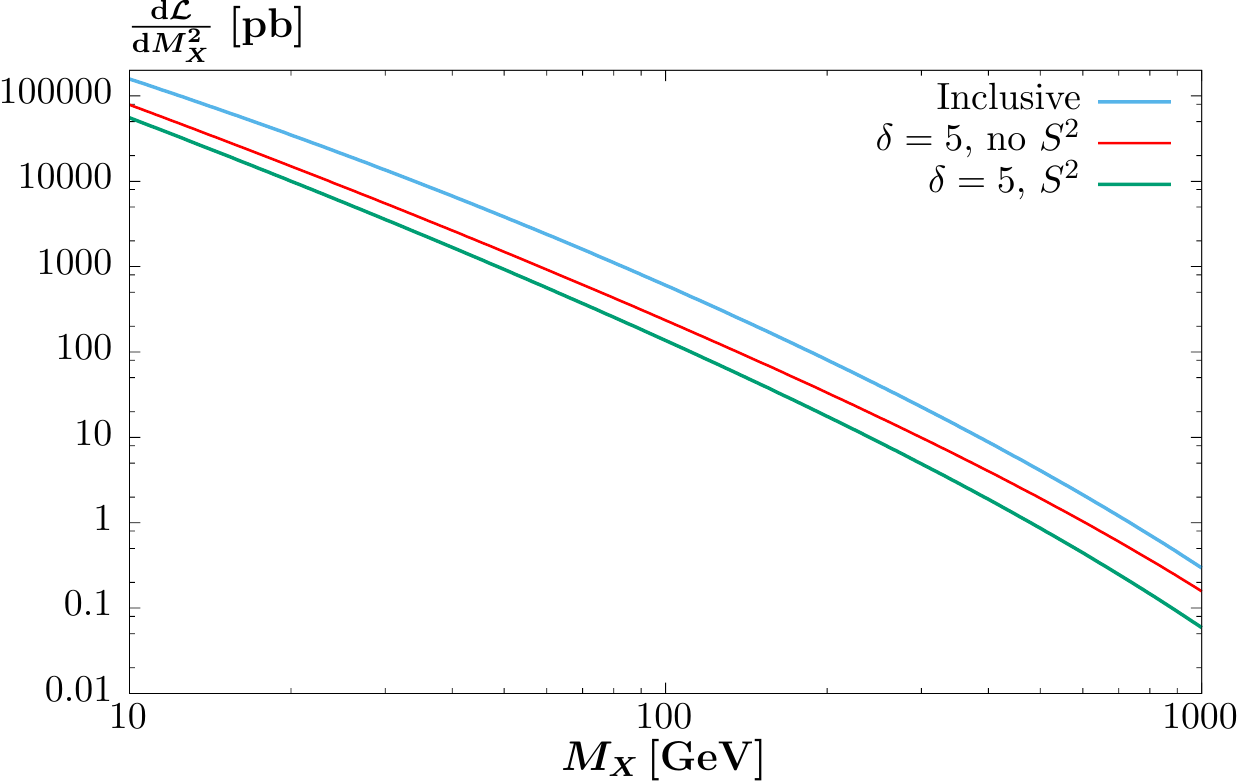}
\includegraphics[scale=0.65]{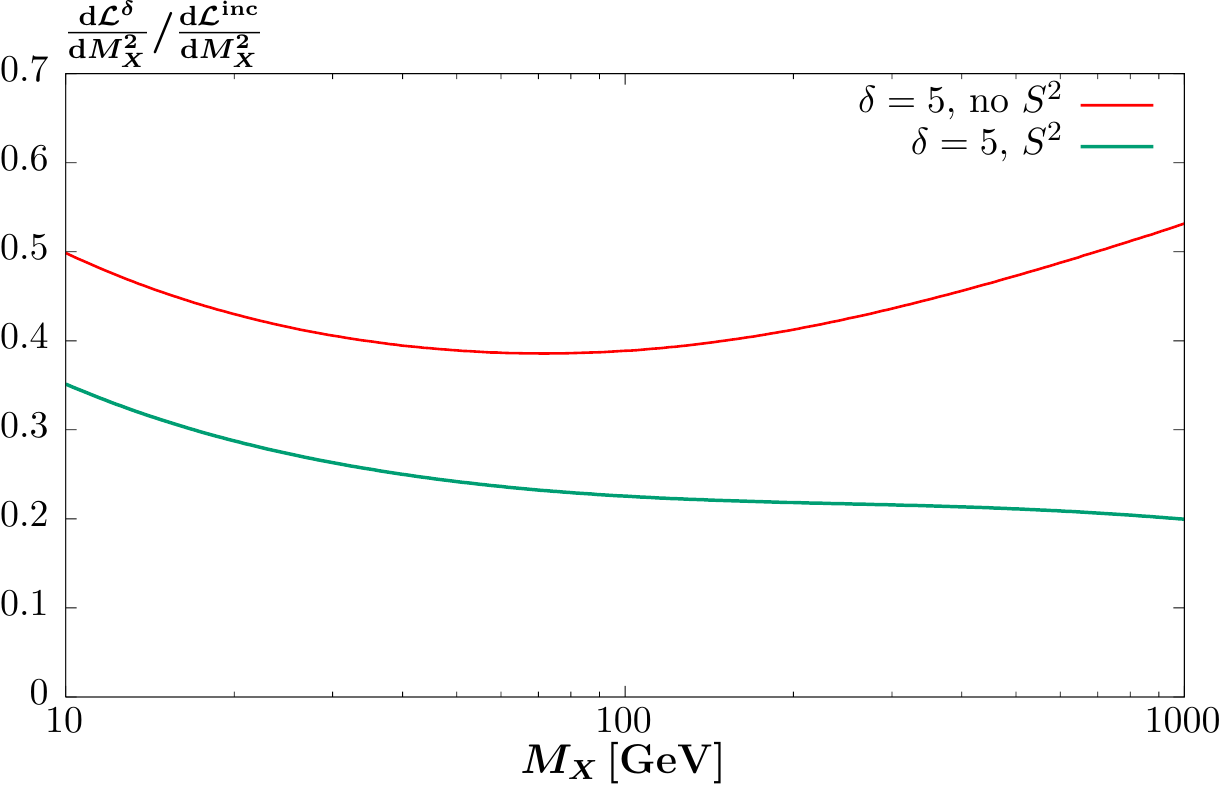}
\caption{$\gamma\gamma$ luminosity at $\sqrt{s}=13$ TeV in the inclusive and semi--exclusive cases, with $\delta=5$ for both protons. For demonstration purposes, the semi--exclusive luminosities are shown both with and without survival effects included. In the left hand figure the absolute luminosities, while in the right hand figure the ratios to the inclusive luminosity are shown.}
\label{fig:lumi}
\end{center}
\end{figure} 

The corresponding average survival factors for all combinations of photon PDF components from each proton are given in Table~\ref{table:surv}. A large range of expected suppression factors is evident, with as anticipated $S^2$ for the lower scale (and hence more peripheral) coherent production process being higher than for the higher scale evolution component. The survival factor for the incoherent component of the input PDF is seen to be particularly small: this is due to the $(1-G_E^2(t))$ factor in (\ref{gamincoh}), which accounts for probability to have no intact proton in the final state, and is therefore peaked towards larger $t$, i.e.  less peripheral interactions, where it is less likely to produce an intact proton.

These results have important implications for the standard factorisation formula
\begin{equation}\label{fact}
\sigma(X)=\int {\rm d}x_1{\rm d}x_2 \,\gamma(x_1,\mu^2)\gamma(x_2,\mu^2)\,\hat{\sigma}(\gamma\gamma\to X)\;,
\end{equation} 
as, using (\ref{photdef1}) and (\ref{inputdef}), we have
\begin{align}\nonumber
\gamma(x_1,\mu^2)\gamma(x_2,\mu^2) &\to
\gamma_{\rm coh}(x_1,\mu^2)\gamma_{\rm coh}(x_2,\mu^2)+\gamma_{\rm incoh}(x_1,\mu^2)\gamma_{\rm incoh}(x_2,\mu^2)\\ \nonumber
&+\gamma_{\rm evol}(x_1,\mu^2;\delta)\gamma_{\rm evol}(x_2,\mu^2;\delta)+\big(\gamma_{\rm coh}(x_1,\mu^2)\gamma_{\rm incoh}(x_2,\mu^2)\\ \nonumber
&+\gamma_{\rm coh}(x_1,\mu^2)\gamma_{\rm evol}(x_2,\mu^2;\delta)+\gamma_{\rm incoh}(x_1,\mu^2)\gamma_{\rm evol}(x_2,\mu^2;\delta)\\&+ 1 \leftrightarrow 2\big)\;.
\end{align}
Crucially, each of these six independent contributions now has a distinct (and in principle, $x$ and/or $\mu$ dependent) survival factor associated with it, and therefore the simple factorisation implied by (\ref{fact}), where the photon flux associated with each proton $i$ can be factorised in terms on an independent PDF $\gamma(x_i,\mu^2)$, no longer holds; instead, this now depends on the state of the other interacting proton, through the influence this has on the survival factor. Physically, this is to be expected, as the survival factor is generated by additional soft proton--proton interactions, which then prevent all of the physics associated with the initial--state photon produced by a given proton being considered independently from the other proton. Such factorisation breaking effects have already been seen in, for example, diffractive production at HERA, where the predictions using the so--called diffractive PDFs are known to dramatically overshoot the data when naively applied in hadron--hadron collisions~\cite{Affolder:2000vb}. 

 \begin{figure}[h]
\begin{center}
\includegraphics[scale=0.65]{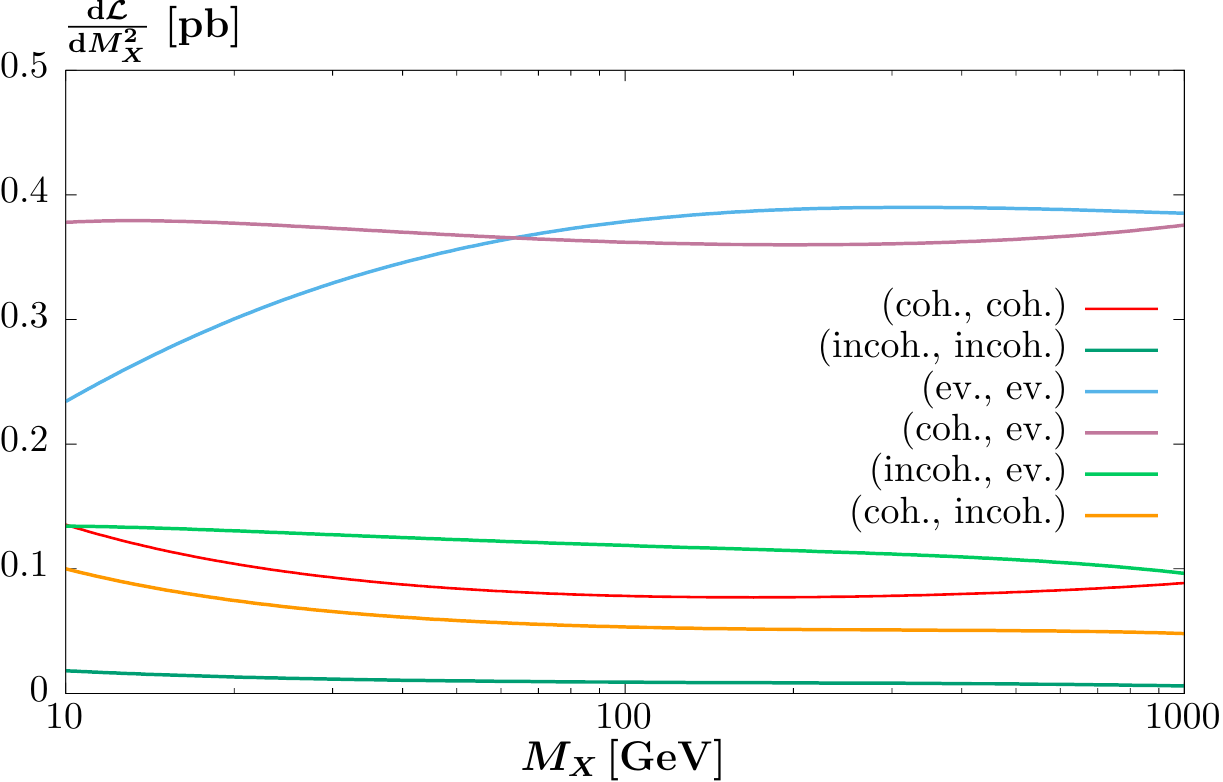}
\includegraphics[scale=0.65]{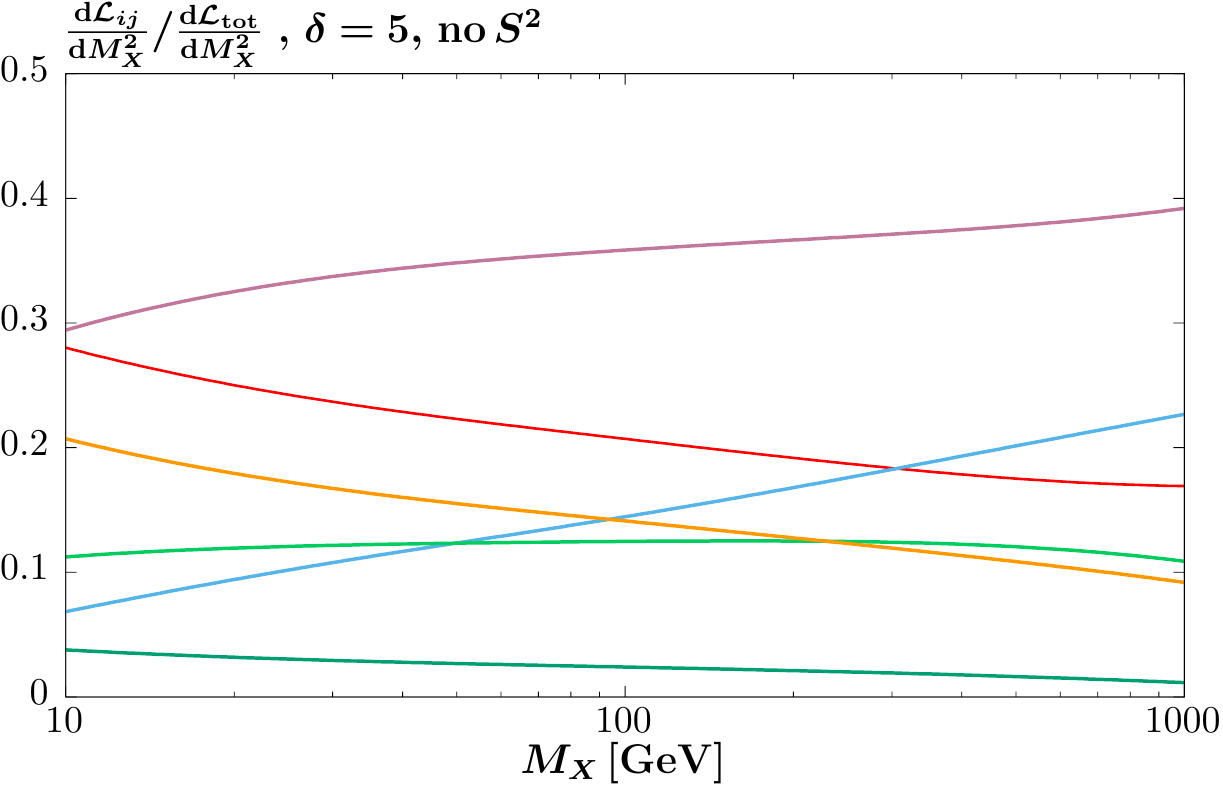}
\includegraphics[scale=0.65]{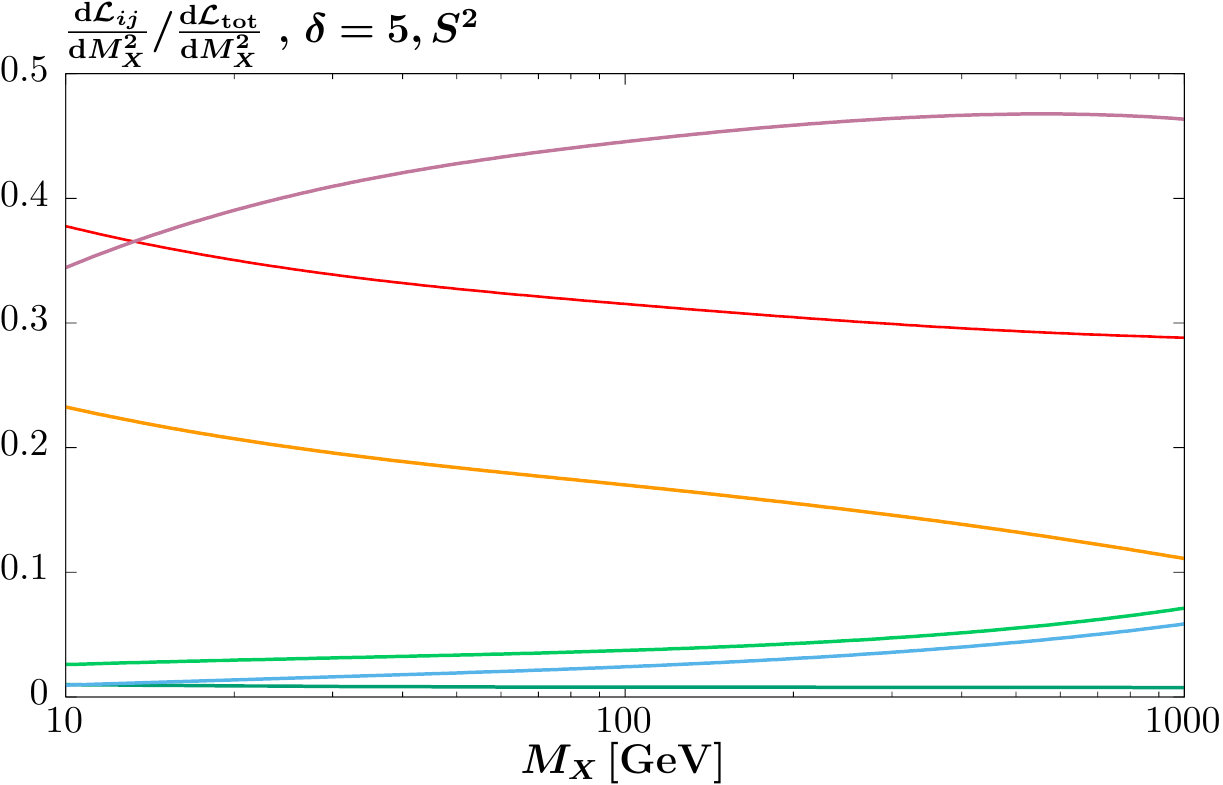}
\caption{Fractional contributions to the photon--photon luminosity at $\sqrt{s}=13$ TeV, defined in (\ref{lumi}), due to the coherent and incoherent input and the evolution components of the photon PDFs, as defined in the text. The upper left panel shows the inclusive case, the upper right the semi--exclusive case, with $\delta=5$ for both protons, but without survival effects included, while the bottom panel is as in the top right, but including survival effects.}
\label{fig:lumic}
\end{center}
\end{figure} 

It is therefore not possible to show equivalent plots to Fig.~\ref{fig:split} demonstrating the impact of survival effects on the individual photon PDF. Instead, we can consider the $\gamma\gamma$ luminosity, given by
\begin{equation}\label{lumi}
\frac{{\rm d}\mathcal{L}}{{\rm d} M_X^2}= \frac{1}{s}\int^1_{\tau} \frac{{\rm d}x_1}{x_1} \,\gamma(x_1,M_X^2)\gamma(\tau/x_1,M_X^2)\;,
\end{equation}
where $\tau=M_X^2/s$ and we take $\mu^2=M_X^2$ as the scale of the PDFs. Such a variable also gives a clearer picture of the suppression we can expect in physical cross sections, as comparing with (\ref{fact}) we have
\begin{equation}
\sigma(X)=\int {\rm d}M_X^2 \,\frac{{\rm d}\mathcal{L}}{{\rm d} M_X^2}\,\hat{\sigma}(\gamma\gamma\to X)\;.
\end{equation}
The photon--photon luminosity at $\sqrt{s}=13$ TeV is shown in Fig.~\ref{fig:lumi} (left), for inclusive and semi--exclusive ($\delta=5$ for both protons) production. For demonstration we show the latter case both with and without survival effects included. We can see that the inclusion of the condition (\ref{dif}) leads to a  factor of $\sim 2$ reduction in the luminosity, roughly consistently with Fig.~\ref{fig:split}, while the inclusion of survival effects leads to a further suppression of  a similar size. That the suppression due to both effects is similar in size is not necessarily to be expected, and indeed for different choices of $\delta$ and/or $\sqrt{s}$, the relative contribution of these effects will differ. It is also interesting to consider how the suppression varies with the central system mass, $M_X$. This is shown in Fig.~\ref{fig:lumi} (right), and in both cases the dependence is seen to relatively mild. The suppression due to introducing the $\delta$ cut decreases at both low and high $M_X$, due to the counteracting effects seen in Fig.~\ref{fig:split}: while increasing $M_X$ leads to a generally larger suppression due to the higher scale at which the PDF is evaluated, this also leads to a larger average $x$ value probed, for which the suppression is less, with similar, but opposite, effects for decreasing $M_X$. Once soft survival effects are included, however, the overall trend is simply decreasing with $M_X$.

A further observable of interest is the relative contribution to the photon--photon luminosity from the different  components of the input photon PDFs, i.e. the coherent, incoherent and evolution defined above, and which roughly speaking correspond to exclusive, low and high--mass diffractive production at the corresponding proton vertex, respectively. This is shown in Fig.~\ref{fig:lumic}, again for the inclusive and semi--exclusive (with/without survival effects) cases. Considering the inclusive case in Fig.~\ref{fig:lumic} (top left), we can see that the incoherent input component is found to be very small, consistently with the results of~\cite{Martin:2014nqa}, while as $M_X$ increases, the relative contribution from the PDF evolution increases, due to the larger evolution length, although this effect is somewhat softened due to the fact that at higher $M_X$ the average $x$ probed increases, where the impact of PDF evolution is less significant. Next, we consider the semi--exclusive case in Fig.~\ref{fig:lumic} (top right). The most dramatic effect is that the relative contribution from the evolution components is  significantly reduced: this is a simple consequence of the fact that the constraint (\ref{dif}) only applies to the PDF evolution. Finally, considering the case with survival effects in Fig.~\ref{fig:lumic} (bottom), we can see that the relative contributions change consistently with the results of Table~\ref{table:surv}. 

The most important consequence of the above results is the suppression seen in the contribution from the evolution components, due both to the impact of (\ref{dif}) on the evolution and the smaller survival factors for this less peripheral interaction. While in the inclusive case the evolution component is larger than the purely coherent one, for the semi--exclusive the result is dramatically different, with the contribution due to the evolution components of both protons being completely negligible. Apart from at the highest $M_X$, it is only the `coherent, evolution' contribution, due to the larger survival factor and smaller impact of (\ref{dif}), which remains significant. Interpreting this result physically, for semi--exclusive production we expect some $M_X$ dependent contribution from exclusive and single proton dissociation, while the double proton dissociation contribution is only noticeable at very high $M_X$.
 
\section{Cross section predictions at the LHC}\label{sec:pred}

\begin{table}
\begin{center}
\begin{tabular}{|c|c|c|c|}
\hline
 & $\sigma^{\mu^+\mu^-}$, $p_\perp^\mu>10$ GeV &$\sigma^{\mu^+\mu^-}$,  $p_\perp^\mu>20$ GeV & $\sigma^{W^+W^- \to l^+ \nu\,l^-\overline{\nu}}$  \\ \hline
$\sigma^{\rm inc}$ [pb]& 12.2 & 2.4 &0.015\\
$\sigma^{\delta=3}/\sigma^{\rm inc.}$& 0.18& 0.16 &0.14\\
$\sigma^{\delta=7}/\sigma^{\rm inc.}$& 0.39& 0.36&0.31 \\
\hline
\end{tabular}
\caption{Cross section predictions for photon--induced muon and $W$ boson pair production at the $\sqrt{s}=13$ TeV LHC. The inclusive and ratio of semi--exclusive to inclusive cross sections are shown, with two choices of emission region $\delta$. In the muon pair case, results for two values of cut on the transverse momenta are shown. For the $W$ boson case, the leptonic decay ($l=e,\mu$) is considered, with lepton transverse momenta $p_\perp^l>25$ GeV, and  missing transverse energy $E_\perp^{\rm miss}>20$ GeV. In all cases the lepton pseudorapidity is required to satisfy  $|\eta|^\mu<2.5$.}
\label{table:cross}
\end{center}
\end{table}

In this section we present a short selection of cross section predictions at the LHC. We will consider for concreteness the cases of muon and $W$ boson pair production at $\sqrt{s}=13$ TeV, although the approach described above may be readily applied to other two--photon induced processes. In principle it would also be possible to apply this approach to the semi--exclusive photoproduction of e.g. vector meson ($J/\psi$, $\Upsilon$...) states, however the situation is then greatly complicated by the presence of the QCD--induced vertex, due in the language of Regge theory to pomeron, rather than photon, exchange; for this reason we do not consider it here.

In Table~\ref{table:cross} we show cross section predictions for photon--induced muon and $W$ boson pair production at the $\sqrt{s}=13$ TeV LHC. We show the inclusive cross section and ratio of semi--exclusive to inclusive cases, with survival effects included and for two values of $\delta$ (applied to both proton sides). These choices are motivated by the experimental situation at the LHC. Considering just the central tracking detector at ATLAS/CMS, we have an uninstrumented region beyond $|\eta|\sim 2.5$, which for $\sqrt{s}=13$ TeV corresponds to roughly $\delta\approx 7$. On the other hand, a much larger rapidity gap may be vetoed on using forward shower counters, which are currently installed at LHCb (the HERSCHEL forward detectors~\cite{Albrow:2014lta}), CMS~\cite{Albrow:2014lta} and ALICE~\cite{Schicker:2014wvk}. Roughly speaking, these extend rapidity coverage out to $|\eta| \lesssim$ 8, corresponding to $\delta\approx 2$ is representative in this case. These can be considered as lower and upper bounds on experimentally realistic values of $\delta$: for other specific experimental configurations, the appropriate value may lie somewhere in between. However, $\delta=2$ represents a relatively small region for non--vetoed emission, which may be sensitive in particular to fluctuations due to fragmentation and hadronisation, see e.g.~\cite{Khoze:2010by}, and for which the transverse momentum of the outgoing quark due to DGLAP evolution satisfying (\ref{dif}) may not be sufficiently high that the simple factorisation (\ref{hft}) holds; for example, if we take a characteristic $z\sim 0.2\, x$ in (\ref{dif}) we have $q_t \lesssim 1.5$ GeV. As well as potentially spoiling this factorisation, this relatively low scale indicates that such a veto may be sensitive to the incoherent input component of the photon PDF, for which the recoil quarks may be produced with sufficiently large $q_t$ to fill the rapidity gap. We thus present results for a somewhat higher value of $\delta=3$, in order to avoid too great a sensitivity to these effects.

We can see from Table~\ref{table:cross} that even for the relatively large $\delta=7$ there is expected to be a significant reduction, by a factor of $\sim 3$, in the predicted cross section relative to the inclusive case; for $\delta=3$ this reduces by a further factor of $\sim 2$, depending on the process. The greater suppression with increasing $M_X$, already seen in Fig.~\ref{fig:lumi}, is also evident. It would be interesting to test these predicted trends, which we recall is a non--trivial result of the formalism described above, in particular that used to model survival effects.

\begin{table}
\begin{center}
\begin{tabular}{|c|c|c|c|}
\cline{3-4}
\multicolumn{1}{}{} & &$\mu^+\mu^-$, $p_\perp^\mu>20$ GeV & $W^+W^- \to \mu^+ \nu\,\mu^-\overline{\nu}$  \\ 
\cline{1-4}
&Exc.& 0.086 &0.079 \\
Inclusive &SD& 0.42&0.41 \\
&DD& 0.50& 0.51\\
\hline
&Exc.& 0.48 & 0.45 \\
$\delta=3$&SD& 0.49& 0.51 \\
&DD& 0.031& 0.045\\
\hline
&Exc.&  0.22 &0.21\\
$\delta=7$&SD& 0.68&0.66 \\
&DD& 0.11  &0.14\\
\hline
\end{tabular}
\caption{Predicted relative contribution of exclusive, single (SD) and double (DD) dissociative components for the same processes as in Table~\ref{table:cross}. Results are rounded to two significant figures, so may not sum to unity precisely.}
\label{table:crossbrea}
\end{center}
\end{table}

In addition, we may investigate the relative contributions of the pure exclusive, single dissociative, and double dissociative components of the photon PDFs to these processes, described in the preceding section. This is shown in Table~\ref{table:crossbrea}: we can see the clear relative enhancement of the pure exclusive component which becomes more pronounced as $\delta$ is decreased. While such a qualitative enhancement in the exclusive component is to be trivially expected when such a rapidity gap veto is applied, the quantitative predictions depend on this particular approach. It is also observed that an increase in the double dissociative component is predicted as $\delta$ increases. 

We end the section with some comments about the theoretical uncertainty in the above predictions. First, it should be emphasised that the coherent component of the input photon PDF, due to purely QED emission from the proton, is theoretically very well understood. Moreover, for purely exclusive processes, i.e. due to coherent emission from both protons, it was found in~\cite{Harland-Lang:2015cta} that the survival factor is almost independent of the specific model taken, as the main model dependence in the evaluation of the soft survival factor lies in the region of small impact parameter $b_t \ll R_p$ (where $R_p$ is the proton radius), whereas for coherent photon emission the average impact parameter is much larger, and so it is relatively insensitive to this region. 

For the incoherent input component, which depends on such issues as the treatment of the quark masses, the uncertainty is larger. We recall that freezing the input $u$ and $d$ quark distributions in (\ref{gamincoh}) at the starting scale, as is done by default, represents an upper limit on the incoherent component of the input distribution. We can instead take the non--relativistic quark model expectations, see~\cite{Martin:2014nqa}, which represent a lower limit on the incoherent component. In this case, for example, the incoherent--coherent component of the $\gamma\gamma$ luminosity in Fig.~\ref{fig:lumic} (bottom) is a factor of $\sim 2$ smaller at lower $M_X$. However, this effect is typically washed out in observable predictions: the inclusive cross sections in Table~\ref{table:cross} are a factor of $\sim 10\%$ smaller, with the ratio of semi--exclusive to inclusive almost unchanged, while the inclusive and $\delta=7$ predictions in Table~\ref{table:crossbrea} are similarly effected at the $\lesssim 10 \%$ level in the case of muon pair production, with the exclusive component being somewhat larger and the single and double dissociation somewhat smaller, while for the higher mass $W^+W^-$ cross section, the effect is even smaller. 
The most sensitive case are the $\delta=3$ predictions for the lower $M_X$ muon pair case, for which the exclusive component is $\sim 20\%$ bigger, the single dissociation $\sim 10\%$ bigger, and the double dissociation is a factor of $\sim 2$ smaller; the effect is similar, but less significant for the $W^+W^-$ cross section. Thus, generally the uncertainty on the incoherent contribution to the input photon PDF affects the cross section prediction at the level of $20\%$ or lower, with the exception of the double dissociation cross section for lower $\delta$, which displays a larger sensitivity. However, this uncertainty may be further reduced by refining the treatment of the input distribution to include higher mass ($\Delta$...) resonances, as well as potentially by constraining it with inclusive measurements that are sensitive to the photon PDF. 

Finally, we must consider the evolution component, where as discussed in the previous section there is some uncertainty in the choice of form factor for the coupling of the photon GPDF to the proton, as in (\ref{hft}). If instead of assuming the same coupling of the GPDF to the GW eigenstates as to the pomeron, we take a universal coupling, this leads to a survival factor of $0.15$ for the evolution--evolution component, i.e. roughly $50\%$ larger than the default value in Table~\ref{table:surv}\footnote{Taking the same form factors as in~\cite{Khoze:2013dha} for the coupling of the pomeron to the GW eigenstates, rather than the proton EM form factor $F_1$, results in a much smaller change.}. However, if we instead consider the coherent--evolution component, then the survival factor is insensitive, at the level of a few percent, to this choice: the interaction is automatically highly peripheral, due to the coherent photon--proton vertex, so that any contribution from the vertex for the evolution component is largely washed out. Thus, while it might appear at first sight that the uncertainty associated with the choice of coupling may be quite large, this is in fact not the case. From Fig.~\ref{fig:lumic} (top right) we can see that the contribution from the evolution--evolution component is small, $\sim 10\%$, even prior to including survival effects, for reasonable values of $M_X$. After including the survival factor, this contribution becomes much smaller still, so that the final cross section prediction will be relatively insensitive to its precise value. In fact, the evolution component enters dominantly through the mixed evolution--coherent term in the $\gamma\gamma$ luminosity, which as discussed above carries little uncertainty due to this choice.

\section{Comparison to CMS semi--exclusive $\mu^+\mu^-$ data}

In~\cite{Chatrchyan:2013akv} the CMS collaboration have presented a measurement of semi--exclusive $W^+W^-$ production via leptonic decays at $\sqrt{s}=7$ TeV. That is, events are selected by demanding a dilepton vertex with no additional associated charged tracks within the tracker acceptance ($\eta<$ 2.4). In this way the contribution from exclusive $\gamma\gamma \to W^+W^-$ is enhanced, and the corresponding cross section measurement can be used to set limits on anomalous quartic gauge couplings. To compare the measured semi--exclusive data with the purely exclusive prediction, a correction factor derived from a larger data sample of $\mu^+\mu^-$ events selected with the same track veto and in the same invariant mass region as the $W^+W^-$ events was applied. Namely, the ratio $F$ of the total number of observed $\mu^+\mu^-$ events, after correcting for the remaining Drell--Yan background, to the prediction found with the \texttt{LPAIR} MC~\cite{Vermaseren:1982cz,Baranov:1991yq}, corresponding to the equivalent photon approximation without any survival factor applied, is taken. Assuming the same factor applies in the $W^+W^-$ case this allows the measured semi--exclusive cross section to be converted to an exclusive one and for corresponding limits on anomalous coupling to be set.

\begin{table}
\begin{center}
\renewcommand\arraystretch{1.2}
\begin{tabular}{|c|c|}
\hline
&$F$\\
\hline
Inclusive&10.9\\
\hline
$\delta=6.5$& 3.6\\
\hline
$\delta=6.5$, $\gamma_{\rm incoh}=0$ & 3.0\\
\hline
\hline
CMS~\cite{Chatrchyan:2013akv}&$3.23\pm 0.53$\\
\hline
\end{tabular}
\caption{Predictions for ratio $F$ of semi--exclusive to exclusive $\mu^+\mu^-$ cross sections, with $M_{\mu\mu}>160$ GeV, $p_\perp^\mu>20$ GeV, $|\eta^\mu|<2.4$ at $\sqrt{s}=7$ TeV, for a rapidity gap veto with $\delta=6.5$, compared to the measurement of~\cite{Chatrchyan:2013akv} which closely corresponds to this scenario. Consistently with~\cite{Chatrchyan:2013akv}, no survival factor is applied for the purely exclusive cross section.}
\label{table:crosscms1}
\renewcommand\arraystretch{1.}
\end{center}
\end{table}

In this paper we are not interested in such limits, but rather on this ratio $F$: assuming that the CMS track requirement corresponds to good approximation to a veto on all additional particle in the $|\eta|<2.4$ region, this is precisely the experimental situation considered in this paper. In particular, for $\sqrt{s}=7$ TeV this corresponds to a rapidity veto with $\delta \approx 6.5$, as defined in Section~\ref{sec:mod}, for $\mu^+\mu^-$ production. In Table~\ref{table:crosscms1} we show our prediction for this scenario\footnote{We use our own prediction for the equivalent photon cross section, rather than \texttt{LPAIR}, although these will coincide closely.} and with the same cuts applied on the final--state muons as in~\cite{Chatrchyan:2013akv}. We can see that the prediction is in excellent agreement with the CMS measurement of $F=3.23\pm 0.53$. Such an encouraging level of agreement, given the various ingredients that enter in the calculation, is not trivial; we recall in addition that the uncertainty on such a result is expected to be at the $\sim 10\%$ level, see the discussion in the preceding section, and hence is of the same order as the experimental uncertainty. We also show for comparison the prediction with the incoherent contribution (\ref{gamincoh}) turned off; this result, which is about $\sim 20\%$ lower, is also completely consistent with the data within current uncertainties. With further data and a reduction in the (dominantly statistical) uncertainty on the measured $F$ it may become possible for the incoherent contribution in our approach to be pinned down, although given that the theoretical uncertainties are of the same order as this $20\%$ difference, this may be challenging.

It is also interesting to consider the predicted ratios $F$ using other available photon PDFs, namely the NNPDF2.3QED~\cite{Ball:2013hta}, CT14QED~\cite{Schmidt:2015zda} and the older MRST2004QED~\cite{Martin:2004dh} sets. This is achieved by applying the same procedure as above, but instead using these sets for the input PDF in (\ref{pdfm}) at the corresponding staring scale $Q_0$ of the set. Unfortunately, for the NNPDF set one added complication is that as no separation is made between the coherent and incoherent inputs, there is a certain amount of freedom in how to treat the survival factor in these cases, which from Table~\ref{table:surv} we can see are generally quite different. As the coherent contribution is expected on general grounds to be dominant, we assume the inputs to be purely coherent when calculating the survival factor, however a more complete treatment would give a somewhat smaller value for $F$; making the unphysical assumption that the input is {\it completely} incoherent, the predicted $F$ is $\sim 30-40\%$ smaller. More realistically, the correct predictions for $F$ could be $\sim 10\%$ smaller than the quoted value. For simplicity, we make the same assumption for the MRST2004QED set, with the results of a more precise treatment being well within the PDF uncertainty range. The CT14QED set is constrained using ZEUS data on isolated photon production in DIS~\cite{Chekanov:2009dq}, and is interpreted in~\cite{Schmidt:2015zda} as being due to the purely inelastic production\footnote{The ZEUS data are selected by requiring that at least on track associated with the proton side is reconstructed. Such a requirement will remove the contribution from the coherent component, and from MC is found to correspond to a constraint of $W_X>5$ GeV on the mass of the produced hadronic system~\cite{forrest}. As at least part of the input incoherent component is expected to fail this extra track requirement, some care may be needed in using these data to constrain the photon PDF, although a study of this is beyond the scope of the current work.}. In the context of our approach, this corresponds to a purely incoherent input component at $Q_0$: we therefore add this to the coherent input (\ref{gamcoh}), with the corresponding survival factors included as described in the previous sections.

In Table~\ref{table:crosscms2} we show predictions for these three PDF sets, with uncertainties calculated as described in the Table caption. In all cases these are completely consistent with the CMS data, within the very large PDF uncertainties. The NNPDF and MRST predictions, which include no explicit coherent input, span a large range above and below the measured value of $F$. On the other hand the CT prediction, for which we have included the coherent contribution explicitly, predicts $F\gtrsim 3$, consistently with our prediction with $\gamma_{\rm incoh}=0$ in Table~\ref{table:crosscms1} (the values are not precisely the same as the evolution is now performed at LO in $\alpha_s$, consistently with~\cite{Schmidt:2015zda}), but extends to significantly higher values of $F$ than in our approach, well beyond the measurement. Given the size of the PDF uncertainties, it is clear that including constraints of this type, even allowing for some conservative theoretical uncertainty, could have a dramatic effect in constraining the photon PDF within these approaches. It is also worth emphasising that the data are in excellent agreement with a dominantly coherent contribution, as calculated in this paper.

\begin{table}
\begin{center}
\renewcommand\arraystretch{1.2}
\begin{tabular}{|c|c|}
\hline
&$F$\\
\hline
CT14&3.1 -- 5.1\\
\hline
NNPDF2.3& $4.0\pm 3.7$\\
\hline
MRST2004 & 1.2 -- 5.3\\
\hline
\hline
CMS~\cite{Chatrchyan:2013akv}&$3.23\pm 0.53$\\
\hline
\end{tabular}
\caption{Predictions for ratio $F$ as in Table~\ref{table:crosscms1}, but with the input photon PDF calculated using the NNPDF2.3QED~\cite{Ball:2013hta} (NLO in QCD, $\alpha_s(M_Z)=0.118$), CT14QED~\cite{Schmidt:2015zda} and the older MRST2004QED~\cite{Martin:2004dh} sets. The MRST2004 range corresponds to the constituent and current quark mass results, the CT14 range to the results with the photon momentum fraction $p_0^\gamma$ between 0 -- 0.14\%, and the NNPDF2.3 uncertainties correspond to a 68\% confidence envelope.
}
\label{table:crosscms2}
\renewcommand\arraystretch{1.}
\end{center}
\end{table}
 
\section{Conclusions}\label{sec:conc}

There are two reasons to include the photon as a parton in the proton. First, in inclusive production the influence of electroweak corrections is increasingly relevant as we enter the era of precision LHC phenomenology, where NNLO QCD calculations are becoming the standard for many processes. In this case it is necessary to introduce the photon PDF in exact analogy to the conventional PDFs of the quarks and gluons, in order to include such corrections consistently. However, the importance  of the photon PDF is also seen in more specific photon--induced processes, such as vector meson photoproduction and $\gamma\gamma$--initiated lepton or $W$ boson pair production. As a rule to select these relatively rare events experimentally some additional cuts, including generally a requirement of large rapidity gaps on either side of the produced object, must be imposed. These cuts affect the incoming photon luminosity and require us to modify the corresponding photon distribution, which no longer corresponds to the usual inclusive one.

In this paper we have considered photon--initiated processes in proton--proton collisions, where one or both protons break up following the interaction. Due to the colour--singlet nature of the photon exchange, these can lead naturally to rapidity gaps in the final--state between the centrally produced system and the proton decay products.  The question, which we have attempted to address, is then what the cross section is for a two--photon initiated process to pass a specific experimental rapidity gap veto. We have shown here how such a veto may be accounted for at leading order by a relatively straightforward adjustment to the usual DGLAP evolution of the photon PDF to account for the kinematic condition implied by the rapidity gap requirement. In addition to this, it is necessary to include the probability for no extra particle production in the gaps due to underlying event activity: the so--called `survival factor'. This object is dependent on soft physics and so introduces an additional model dependence to the calculation. However, we have shown explicitly that provided the rapidity region for allowed emission is not too restrictive, then the uncertainty due to this is quite limited. This is in fact not surprising, and is a result of a more general point: it is the relatively simple nature of the photon emission (in particular, the lack of colour exchange induced) which allows for the rapidity gap veto to be included in the DGLAP evolution in a straightforward way, while the peripheral nature of the photon exchange leads to the uncertainty due to the survival factor in the observable cross section generally being small.

As a result of the effects described above we expect the `effective' $\gamma\gamma$ luminosity and hence predicted cross section to be reduced  when an experimentally realistic rapidity gap veto is applied to both sides of the produced state, with the impact of vetoing on quark emission in the PDF DGLAP evolution and the survival factor being comparable in size. The level of suppression depends on the exact gap size, as well as the hard process and event selection criteria; some representative predictions for realistic experimental scenarios have been presented in Table~\ref{table:cross}, where the overall suppression is seen to lie in the range $\sim 2-7$. In Table~\ref{table:crossbrea} the breakdown between the fraction of purely exclusive events and those with one or both protons dissociating for a selection of scenarios has been given; while demanding rapidity gaps will naturally lead to an enhancement of exclusive events, the precise predictions for these fractions and their dependence on the gap size is a non--trivial result of the theoretical approach, in particular soft survival effects, which do not enter uniformly and are expected to suppress the dissociative contributions beyond naive estimates. Depending on the gap size, we expect the fraction of exclusive events to be $\sim 20 - 50\,\%$, with the remaining events being principally due to single proton dissociation, while the contribution from double dissociation is below $\sim 15\%$, and may be significantly lower.  In all cases, we have demonstrated that it is crucial to fully account for the experimental cuts, and in particular the rapidity gap veto, when comparing data to theoretical predictions for photon--initiated production;  it is not sufficient to use the conventional inclusive photon PDF. 

Such rapidity gap events are of much phenomenological interest at the LHC, with a range of measurement possibilities,  using just the central tracking detector at ATLAS/CMS, or across the wider rapidity region allowed by the forward shower counters currently installed at LHCb (the HERSCHEL forward detectors~\cite{Albrow:2014lta}), CMS~\cite{Albrow:2014lta} and ALICE~\cite{Schicker:2014wvk}. We have therefore presented a selection of cross section predictions for semi--exclusive photon--initiated lepton and $W$ boson pair production at the $\sqrt{s}=13$ TeV LHC, with rapidity gap veto regions relevant to both these scenarios. The cross sections are quite large, and the relatively simple formalism outlined in this paper leads to some clear predictions for such observables as the ratios of the single and double proton dissociative contributions to the total cross section, and their dependence on the central system mass $M_X$ and rapidity gap veto size. By measuring these at the LHC, a better understanding of the photon PDF and models of soft physics may be possible.

\section*{Acknowledgements}

The work of MGR  was supported by the RSCF grant 14-22-00281. VAK thanks the Leverhulme Trust for an Emeritus Fellowship.  LHL thanks the Science and Technology Facilities Council (STFC) for support via the grant award ST/L000377/1. MGR thanks the IPPP at the University of Durham for hospitality.

\bibliography{references}{}

\begin{thebibliography}{10}

\bibitem{Martin:2004dh}
A.~D. Martin, R.~G. Roberts, W.~J. Stirling, and R.~S. Thorne,
\newblock Eur. Phys. J. {\bf C39}, 155 (2005), hep-ph/0411040.

\bibitem{Ball:2013hta}
NNPDF, R.~D. Ball {\em et~al.},
\newblock Nucl. Phys. {\bf B877}, 290 (2013), 1308.0598.

\bibitem{Schmidt:2015zda}
C.~Schmidt, J.~Pumplin, D.~Stump, and C.~P. Yuan,
\newblock (2015), 1509.02905.

\bibitem{Kepka:2008yx}
O.~Kepka and C.~Royon,
\newblock Phys.Rev. {\bf D78}, 073005 (2008), 0808.0322.

\bibitem{Chapon:2009hh}
E.~Chapon, C.~Royon, and O.~Kepka,
\newblock Phys.Rev. {\bf D81}, 074003 (2010), 0912.5161.

\bibitem{Royon:2015coa}
C.~Royon and M.~Saimpert,
\newblock EPJ Web Conf. {\bf 90}, 06001 (2015).

\bibitem{Aad:2015bwa}
ATLAS Collaboration, G.~Aad {\em et~al.},
\newblock (2015), 1506.07098.

\bibitem{Harland-Lang:2015cta}
L.~A. Harland-Lang, V.~A. Khoze, and M.~G. Ryskin,
\newblock (2015), 1508.02718.

\bibitem{Khoze:2000db}
V.~A. Khoze, A.~D. Martin, R.~Orava, and M.~G. Ryskin,
\newblock Eur.Phys.J. {\bf C19}, 313 (2001), hep-ph/0010163.

\bibitem{ATLAS:2015exc}
ATLAS-CONF-2015-081.

\bibitem{CMS:2015dxe}
CMS-PAS-EXO-15-004.

\bibitem{Fichet:2015vvy}
S.~Fichet, G.~von Gersdorff, and C.~Royon,
\newblock (2015), 1512.05751.

\bibitem{Csaki:2015vek}
C.~Csaki, J.~Hubisz, and J.~Terning,
\newblock (2015), 1512.05776.

\bibitem{Fichet:2016pvq}
S.~Fichet, G.~von Gersdorff, and C.~Royon,
\newblock (2016), 1601.01712.

\bibitem{Csaki:2016raa}
C.~Csaki, J.~Hubisz, S.~Lombardo, and J.~Terning,
\newblock (2016), 1601.00638.

\bibitem{yp}
K. Akiba et al, CERN-PH-LPCC-2015-001, available at {\tt
  http://www-d0.fnal.gov/Run2Physics/qcd/loi\_atlas/fpwg\_yellow\_report.pdf}.

\bibitem{Albrow:2014lta}
M.~Albrow, P.~Collins, and A.~Penzo,
\newblock Int.J.Mod.Phys. {\bf A29}, 1446018 (2014).

\bibitem{Schicker:2014wvk}
R.~Schicker,
\newblock Int.J.Mod.Phys. {\bf A29}, 1446015 (2014), 1411.1283.

\bibitem{CERN-LHCC-2011-012}
CERN Report No. CERN-LHCC-2011-012. LHCC-I-020, 2011 (unpublished).

\bibitem{Albrow:1753795}
M.~Albrow {\em et~al.},
\newblock CERN Report No. CERN-LHCC-2014-021. TOTEM-TDR-003. CMS-TDR-13, 2014
  (unpublished).

\bibitem{Royon:2015tfa}
C.~Royon and N.~Cartiglia,
\newblock Int.J.Mod.Phys. A29 (2014) 28, 1446017  (2015), 1503.04632.

\bibitem{Chatrchyan:2013akv}
CMS collaboration, S.~Chatrchyan {\em et~al.},
\newblock JHEP {\bf 07}, 116 (2013), 1305.5596.

\bibitem{Khoze:2014aca}
V.~A. Khoze, A.~D. Martin, and M.~G. Ryskin,
\newblock Int.J.Mod.Phys. {\bf A30}, 1542004 (2015), 1402.2778.

\bibitem{Gotsman:2014pwa}
E.~Gotsman, E.~Levin, and U.~Maor,
\newblock Int.J.Mod.Phys. {\bf A30}, 1542005 (2015), 1403.4531.

\bibitem{deFlorian:2015ujt}
D.~de~Florian, G.~F.~R. Sborlini, and G.~Rodrigo,
\newblock (2015), 1512.00612.

\bibitem{Martin:2014nqa}
A.~D. Martin and M.~G. Ryskin,
\newblock Eur. Phys. J. {\bf C74}, 3040 (2014), 1406.2118.

\bibitem{Bertone:2013vaa}
V.~Bertone, S.~Carrazza, and J.~Rojo,
\newblock Comput. Phys. Commun. {\bf 185}, 1647 (2014), 1310.1394.

\bibitem{Harland-Lang:2014zoa}
L.~A. Harland-Lang, A.~D. Martin, P.~Motylinski, and R.~S. Thorne,
\newblock (2014), 1412.3989.

\bibitem{Harland-Lang:2014lxa}
L.~A. Harland-Lang, V.~A. Khoze, M.~G. Ryskin, and W.~Stirling,
\newblock Int.J.Mod.Phys. {\bf A29}, 1430031 (2014), 1405.0018.

\bibitem{Ryskin:2011qe}
M.~G. Ryskin, A.~D. Martin, and V.~A. Khoze,
\newblock Eur.Phys.J. {\bf C71}, 1617 (2011), 1102.2844.

\bibitem{Ryskin:2009tk}
M.~G. Ryskin, A.~D. Martin, and V.~A. Khoze,
\newblock Eur.Phys.J. {\bf C60}, 265 (2009), 0812.2413.

\bibitem{Martin:2009ku}
A.~D. Martin, M.~G. Ryskin, and V.~A. Khoze,
\newblock Acta Phys.Polon. {\bf B40}, 1841 (2009), 0903.2980.

\bibitem{Antchev:2013gaa}
TOTEM collaboration, G.~Antchev {\em et~al.},
\newblock Europhys. Lett. {\bf 101}, 21002 (2013).

\bibitem{Diehl:2003ny}
M.~Diehl,
\newblock Phys.Rept. {\bf 388}, 41 (2003), hep-ph/0307382,
\newblock Habilitation thesis.

\bibitem{Belitsky:2005qn}
A.~V. Belitsky and A.~V. Radyushkin,
\newblock Phys.Rept. {\bf 418}, 1 (2005), hep-ph/0504030.

\bibitem{Ji:1996ek}
X.-D. Ji,
\newblock Phys. Rev. Lett. {\bf 78}, 610 (1997), hep-ph/9603249.

\bibitem{Khoze:2013dha}
V.~A. Khoze, A.~D. Martin, and M.~G. Ryskin,
\newblock Eur.Phys.J. {\bf C73}, 2503 (2013), 1306.2149.

\bibitem{Good:1960ba}
M.~L. Good and W.~D. Walker,
\newblock Phys. Rev. {\bf 120}, 1857 (1960).

\bibitem{Affolder:2000vb}
CDF collaboration, T.~Affolder {\em et~al.},
\newblock Phys. Rev. Lett. {\bf 84}, 5043 (2000).

\bibitem{Khoze:2010by}
V.~A. Khoze, F.~Krauss, A.~D. Martin, M.~G. Ryskin, and K.~C. Zapp,
\newblock Eur. Phys. J. {\bf C69}, 85 (2010), 1005.4839.

\bibitem{Vermaseren:1982cz}
J.~Vermaseren,
\newblock Nucl.Phys. {\bf B229}, 347 (1983).

\bibitem{Baranov:1991yq}
S.~P. Baranov, O.~Duenger, H.~Shooshtari, and J.~A.~M. Vermaseren,
\newblock {Hamburg 1991, Proceedings, Physics at HERA, vol. 3, 1478-1482. (see
  HIGH ENERGY PHYSICS INDEX 30 (1992) No. 12988)}  (1991).

\bibitem{Chekanov:2009dq}
ZEUS, S.~Chekanov {\em et~al.},
\newblock Phys. Lett. {\bf B687}, 16 (2010), 0909.4223.

\bibitem{forrest}
M. Forrest. Isolated photon production in deep inelastic scattering at HERA -
  2010. PhD Thesis.

\end{thebibliography}
\bibliographystyle{h-physrev}

\end{document}